# NEO-PGA: Nonvolatile electro-optically programmable gate array

Rui Chen[1,2,*], Andrew Tang[1], Jayita Dutta[1], Virat Tara[1], Julian Ye[3], Zhuoran Fang[1], Arka Majumdar[1,3,*]

[1]Department of Electrical and Computer Engineering, University of Washington, Seattle, WA 98195, USA

[2]Department of Materials Science and Engineering, Massachusetts Institute of Technology, MA 02139, USA

[3]Department of Physics, University of Washington, Seattle, WA 98195, USA

*Email: charey@mit.edu and arka@uw.edu

# Abstract

Programmable photonic integrated circuits (PICs) offer a unique opportunity to create a flexible platform, akin to electronic field programmable gate array (FPGA). These photonic PGAs can implement versatile functionalities for applications ranging from optical interconnects to microwave photonics. However, state-of-the-art programmable photonics relies predominantly on volatile thermo-optic tuning, which suffers from high static power consumption, large footprints, and thermal crosstalk. All these dramatically limit the gate density and pose a fundamental limit to the scalability. Chalcogenide-based phase-change materials (PCMs) offer a superior alternative due to their nonvolatility and substantial optical contrast, though challenges such as optical loss, and bit precision severely limited their application in large-scale PICs. Here, we demonstrate precise, multi-bit, low-loss tuning of the emerging PCM $Sb_2Se_3$ using a closed-loop, "program-and-verify" method. Electrically reconfigurable PCM-integrated silicon photonic gates are implemented on a 300 mm silicon photonic platform, using circulating and forward Mach–Zehnder interferometer (MZI) meshes. In the circulating mesh, we realize broadband optical switching fabrics and high-Q coupled resonators with unprecedented local control of coupling rates, which further enable exploration of coupled-cavity systems. The forward mesh supports self-configurable MZIs that sort two orthogonal beams to different ports. These results showcase a new type of scalable photonic PGA enabled by PCMs, offering a pathway toward general-purpose, on-chip programmable photonic systems.

# Introduction

Photonic integrated circuits (PICs) are critical for many applications, including optical transceivers[1], on-chip spectrometers[2–5], radiofrequency (RF) filters[6,7] and photonic information processing (both classical and quantum)[8–10]. In most cases, each application requires a separate PIC design, akin to the electronic application-specific integrated circuits (ASICs), dramatically elongating the prototyping cycle and increasing cost. As an alternative, programmable PICs[11,12] have attracted significant attention due to their multifunctional capabilities, offering a more versatile platform that enables an optical analogue to electronic field-programmable gate arrays (FPGAs).

These photonic PGAs are usually composed of tunable components such as phase shifters and resonators. One commonly used unit cell is the Mach-Zehnder Interferometer (MZI) with two phase shifters, which allow independent tuning of both the amplitude and the phase[12]. Programmable PICs can be reconfigured into vastly different architectures to implement diverse functionalities. A typical architecture consists of MZI gates arranged in closed meshes[13], such as rectangular[7] or hexagonal[11] arrays. By programming these gates, we can make light propagate in a forward direction or circulate in a closed loop, creating resonators. Such ability to arbitrarily route light on-chip allows the same gate array to be used as tunable RF filters[7,14], high-resolution RF beamforming structures[15], arbitrary unitary transformation[16,17] for quantum optical information processing[18], on-chip spectrometer[2], beam steering devices[19], coherent beam combiner[20,21] or even brain neural probes[22]. Recent experiments have also verified the utility of such PGAs in determining the optimal communication channels for an unknown environment perturbation[23]. As such, photonic PGAs can help consolidate a historically fragmented photonic industry focusing on diverse applications.

Unfortunately, existing photonic PGAs employ weak and volatile thermo-optic phase shifters. This incurs a substantial amount of static power (~10 mW/$\pi$[24]), large device footprints (> 50 $\mu m$[25]) and severe thermal crosstalk (requiring ~100 $\mu m$ spacing between shifters). As PICs are scaling towards thousands of components per die[26], the power consumption of thermo-optic phase shifters can easily amount to several Watts. A truly "set-and-forget" functionality can revolutionize the field of optical PGAs. Chalcogenide-based nonvolatile phase-change materials (PCMs) offer a promising solution to all these problems[27–29]. PCMs exhibit two optically distinct, ambient stable, and reversibly switchable micro-structural phases (amorphous and crystalline), which render nonvolatile tuning of both amplitude and phase of light. To change PCMs from amorphous to crystalline state, they are heated above the crystallization temperature. To reamorphize, PCMs are first melted by heating above the melting point and rapidly quenched into the metastable amorphous phase. This threshold-driven switching behavior enables highly localized (~1$\mu m$) and crosstalk-free tuning[30] - capabilities that are unattainable with any existing tuning methods, including thermo-optic/ electro-optic effects or liquid crystal[31,32]. Previously, optical information was primarily encoded in the loss of PCMs, excluding them from phase-only MZI mesh circuits and limiting their applications to relatively small-scale systems[33–36]. Moreover, electrical programming of PCMs was deemed stochastic[37], which compromised the precision of optical levels. Although emerging wide bandgap PCMs, such as $Sb_2S_3$[38] and $Sb_2Se_3$[39], show promises to reduce the loss in the short-wave infrared wavelength range, they are demonstrated only at a single device level, such as a single phase shifter[40–45] and tunable directional coupler[41,43,46]. To date, no system-level demonstration of PCM-based, phase-only programmable PICs has been reported. As such, the notion that electrically switched PCMs cannot provide accurate bidirectional tuning and hence are ill-suited for programmable PICs is prevalent in the community.

In this paper, we demonstrate electrically programmable PGAs made of PCM-integrated MZI meshes. Leveraging a "program-and-verify" algorithm with closed-loop feedback[47], we show highly accurate (within ~0.1% error) tuning of optical power. This demonstration relied on bidirectional and progressive tuning of PCM. Our method utilizes a close-loop, automated electrical pulse sequence to drive the PCM to a desired state and thus programming a MZI with specified transmission. To establish the versatility of our PGAs, we demonstrated two classes of systems, including both circulating and forward-only PICs. In the first class of PIC, we report broadband optical signal routing (most of the ports have crosstalk < -25 dB) and narrowband ring resonators with programmable coupling (intrinsic quality factor ~226,000). In the same system, we reconfigured it into a decoupled dual-ring system showing independent tuning of one ring without perturbing another placed in close proximity. This validates the highly local and crosstalk-free tuning capability of PCMs. Furthermore, by controlling coupling between two rings, we experimentally studied coupled-cavity systems and realized transition from strong to weak coupling regime. In the second class of PIC, we reported MZI-based forward meshes and realized a coherent mode-converters (crosstalk ~ -12 dB) via recently proposed self-configurable PIC[21]. Through these demonstrations, we challenge the prevailing notion that PCMs are unsuitable for programmable PICs due to their inherent stochasticity, a limitation that is effectively mitigated through the "program-and-verify" algorithm. Our results open a new Nonvolatile Electro-Optically Programmable Gate Array (NEO-PGA), that truly functions as an on-chip analogue of the free-space kinematic mounts, which are reconfigured to align a macroscopic (on-chip) optical setup and then locked to avoid any excess power consumption.

## Results

Using a scalable back-end-of-line (BEOL) integration method, we heterogeneously integrated PCMs onto high volume manufactured silicon photonic dies[43]. Only one electron-beam lithography step is performed for $Sb_2Se_3$ patterning (fabrication resolution of ~100 nm), which can be readily replaced with advanced deep ultra-violet photolithography in the future. The detailed fabrication process can be found in Method and Supplementary Section 1.

### Bi-directional electrical programming of a non-volatile Mach-Zehnder unit

To illustrate the key idea of closed-loop programming and the "program-and-verify" algorithm, we first show the function of a single programmable unit in Figure 1a, *i.e.* a balanced MZI consisting of two 50:50 multimode interferometers (MMIs) and a non-volatile $Sb_2Se_3$ phase shifter on each arm. The MMIs are from Intel's Process Design Kits (PDKs), while the customized non-volatile phase shifter consists of a thin film of ~10-nm $Sb_2Se_3$ on the top surface of a 400-nm-wide, 65-$\mu m$-long silicon ridge waveguide with a total and slab height of 300 nm and 100 nm, respectively. The silicon slab is doped to form a P++-doped/Intrinsic/N++-doped (PIN) microheater 500 nm away from the waveguide edges, similar to our previous demonstrations[43]. We use PIN-diode silicon heaters instead of N++-N-N++ resistive heaters because PIN diodes introduce almost zero excess static loss. Although they are still lossy when passing current (during the switching event), such free-carrier dispersion loss is transient and returns to zero once the pulse is terminated, and the device is switched to the new nonvolatile state. Different pulses (power and width) are employed to switch the $Sb_2Se_3$ thin films to amorphous or crystalline phases. A schematic drawing of our $Sb_2Se_3$ tunable MZI and a cross-sectional view of the phase shifter are shown in Figure 1a. The optical microscope images of the fabricated devices are shown in Figure 1b.

Scalable programmable photonics require low insertion loss for each component. The wide bandgap PCM $Sb_2Se_3$ exhibits extremely low loss[39] (Supplementary Section 2 for measured complex refractive index), which is otherwise unachievable with conventional PCMs such as GST[48,49]. Simulation shows a $\pi$-phase shift length of 65 $\mu m$ with 10 nm of $Sb_2Se_3$, and a loss of ~0.03 dB/$\pi$. The phase shift and loss are further validated by an array of microring resonators, which are loaded with $Sb_2Se_3$ of varying lengths (Supplementary Section 3). By fitting the measured ring resonances to Lorentzian line shapes, we extract a phase shift of 0.013 $\pi/\mu m$ and crystalline phase loss of only 0.037 dB/$\pi$.

Besides the low insertion loss, programmable PICs also require sufficient tuning ranges. Figure 1c shows the measured spectrum of a balanced MZI, with a switching condition for amorphization (crystallization) of 7.9 V, 400 ns (2.9 V, 5 ms). The experiment is highly repeatable with minimal standard deviations among 5 experiment cycles, indicated by the shaded region. We note that the reversible phase tuning range in these experiments is around 0.23$\pi$, significantly smaller than the simulation or our switching results from rapid thermal annealing. This can be attributed to inadequate material deposition or capping process, which causes a large portion of the material's switching difficult to switch. In the experiments reported in this paper, large phase tuning led to a reduced phase shift, which can be improved with better material processing.

The last prerequisite for programmable PICs is quasi-continuous and bidirectional tuning, which allows the system to correct an over-tuned configuration based on closed-loop feedback. Essentially a bidirectional tuning ensures that, if we overshoot or undershoot our desired level, we can modify our current level slightly, rather than, going back to the initial state and start the programming again. To show that $Sb_2Se_3$ based silicon phase shifters can satisfy this, we applied incremental voltages from 6.8 V to 7.1 V with a fixed pulse duration of 400 ns, thereby setting the device to different partial amorphization states. We emphasize that given this is a small change, we do observe reversible tuning in these experiments. The blue curve in Figure 1d shows the temporal transmission at 1340 nm while these partial amorphization pulses are applied. Depending on the pulse voltage plotted in the bottom figure, multiple levels can be accessed. Conversely, partial crystallization was also achieved by gradually increasing the crystallization voltage from 1 V to 3 V with a pulse duration of 5 ms, shown by the orange curves in Figure 1e. Figure 1d and Figure 1e prove that the quasi-continuous tuning is bidirectional, providing the capability of performing the "program-and-verify" algorithm.

As an illustration, we first accomplish a simple yet crucial task with a single component – accurate tuning of MZI's transmission level by monitoring the cross output and changing the pulses applied to the phase shifter. Figure 1f demonstrates an automated programming of an MZI targeting a transmission of -15 dB with less than 0.1% error. We note that, without external feedback, we cannot achieve such a low error, and we had an error of ~3.1% (defined by standard deviation divided by mean value) among three cycles (Supplementary Section 4). The switching voltages are determined by the power readout after each pulse (Supplementary

Section 5 for algorithm for automated control). In amor-2 sequence, the slight increase in the optical transmission at the beginning is attributed to unintended slight crystallization with a low voltage. Despite the transmission evolution not being identical in different cycles (see Supplementary Section 6), the target transmission is achieved in the end with high precision, validating the importance of the "program-and-verify" algorithm. In principle, this "program-and-verify" method works for arbitrary desired output power (Supplementary Section 7). This high precision tuning of PCMs defies the previous belief that PCMs cannot be used for deterministic tuning due to their stochastic multi-bit operation.

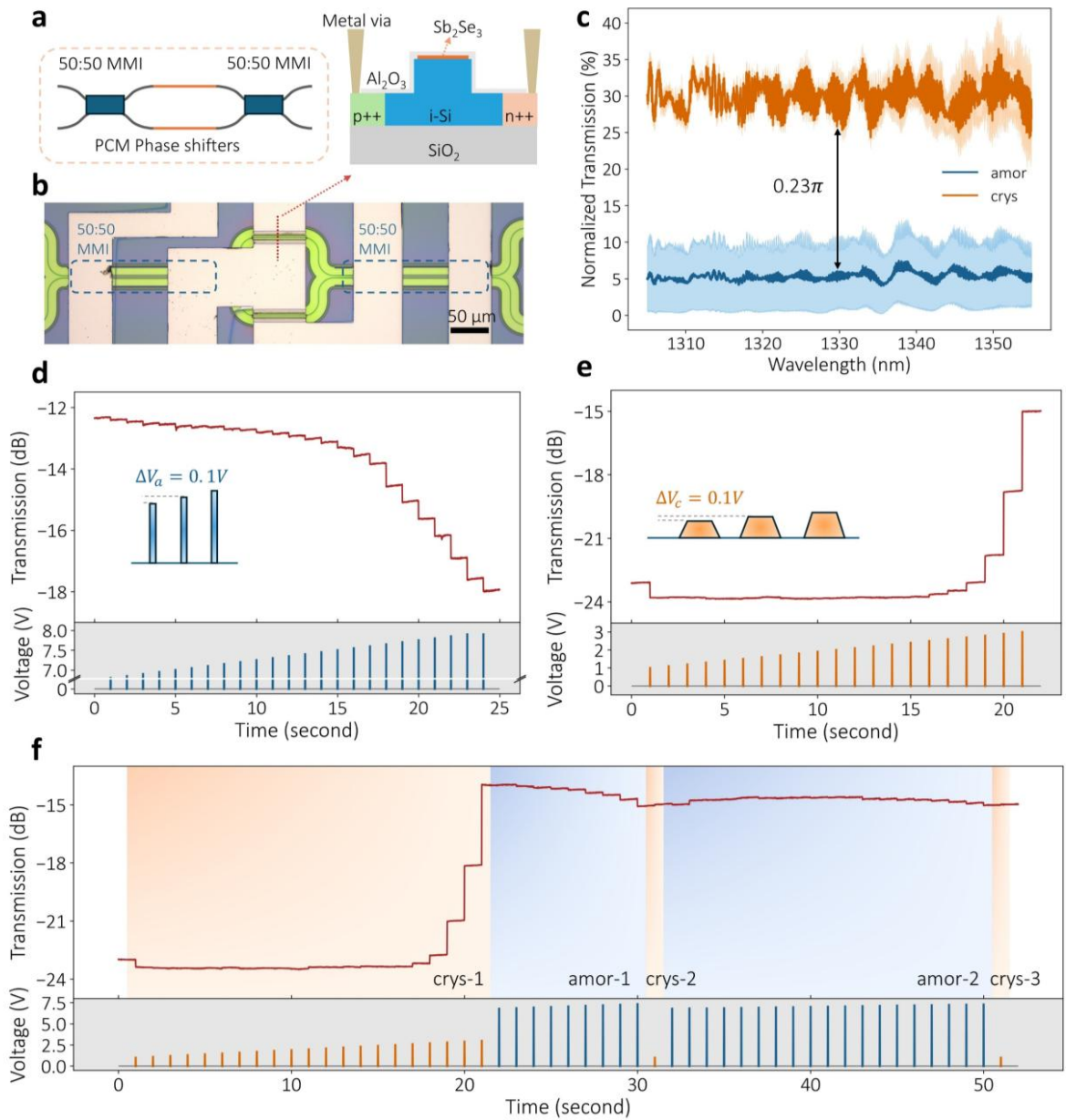

Figure 1: **The non-volatile, automated programming of Sb$_2$Se$_3$-clad Mach-Zehnder Interferometer (MZI) unit. a.** Schematic and the cross-sectional view of the Sb$_2$Se$_3$-Si phase shifter in the MZI unit. **b.**

An optical micrograph of a fabricated MZI test unit. The 50:50 MMIs are boxed in blue dashed line. The red dotted line annotates the cross-section in **a**. **c.** Averaged bar-port spectra of the MZI unit for 5 cycles, with shaded region indicating the standard deviation. **d. e.** Time trace data showing the device can access multiple levels by both **d** partial amorphization and **e** partial crystallization pulses, respectively. The bottom figures show the corresponding pulse voltage in the same time frame (due to long time axis, the different pulse durations for amorphization and crystallization are not shown). **f.** Automated tuning of a balanced MZI to a target bar-port transmission of -15 dB with ~0.1% error. The instruments are computer-controlled using a "program-and-verify" algorithm. Each pulse is separated by one second to allow thermal stabilization although this time separation can be readily decreased to ~1 ms.

## Circulating Mach-Zehnder mesh structure for optical circuit switch

Leveraging the $Sb_2Se_3$-enabled MZI units, we next demonstrate zero-static-power tunning of circulating MZI meshes, which are commonly used for microwave filtering[7], microwave beam forming[15] and optical signal routing[14]. Figure 2a shows an optical micrograph of $1 \times 2$ rectangular MZI meshes actively programmable by $Sb_2Se_3$. We chose to employ the rectangular shape instead of triangular and hexagonal topologies, because of its simpler architecture and more intuitive programming algorithms. We first use these simple $1 \times 2$ rectangular MZI meshes to implement an on-chip $4 \times 4$ optical circuit switching (OCS) fabric. Programming each MZI to fully cross or bar states, the OCS can guide a specific input pattern (denoted as $I_{1234}$) to different output pattern (denoted as $O_{y_1 y_2 y_3 y_4}$; here $y_i$ represents the output port, from where the light input at $i^{th}$ port comes out). Thus routing $I_{1234}$ to $O_{y_1 y_2 y_3 y_4}$ signifies, light input at ports 1,2,3 and 4 will be routed to the output ports $y_1, y_2, y_3$ and $y_4$ respectively. In our programmable PIC, we often route one input (from port n) to one output (port m), and that path is denoted as $I_n O_m$. Two example configurations are shown in Figure 2b and Figure 2c, where the input light is guided to output ports $O_{4231}$ and $O_{4213}$, respectively.

It is noteworthy that our current OCS is not non-blocking, because it can only realize 16 (Supplementary Section 8 for all possible outputs) out of the total $4! = 24$ output configurations. However, a non-blocking OCS can be readily achieved with a larger mesh, such as $1 \times 4$ circulating mesh (Supplementary Section 8).

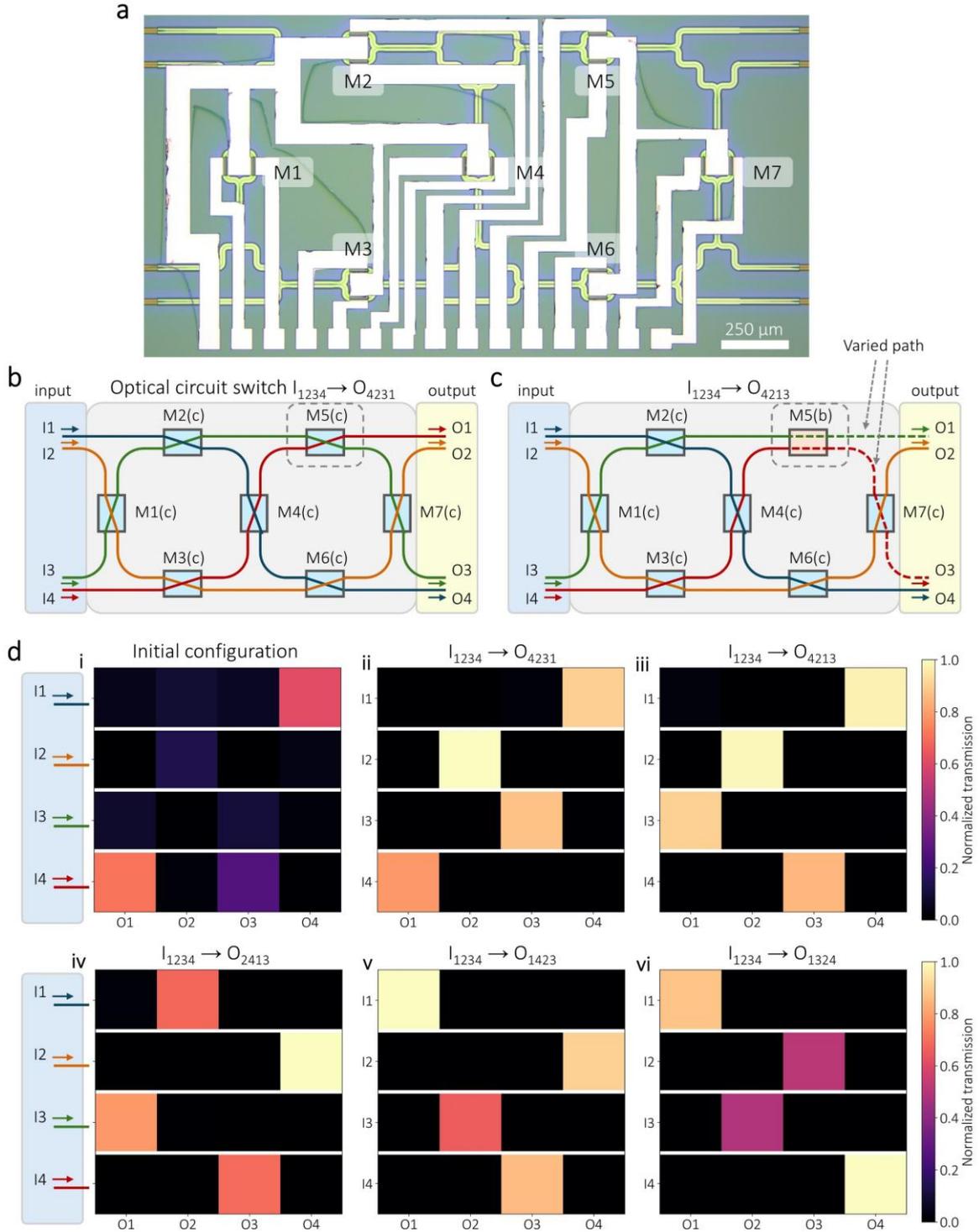

Figure 2: **Rectangle meshes implementing an optical circuit switching fabric. a.** Optical micrograph of the 1 × 2 circulating MZI meshes (Scale bar: 250 μm). Each MZIs are annotated corresponding to the schematics in **b** and **c**. **b. c.** A schematic of a configuration where light is guided from input ports $I_{1234}$ to output ports $O_{4231}$ and $O_{4213}$, respectively. $M_i$'s denote different MZI switches and "(c)" and "(b)" denote a cross or a bar state. Red, green, blue and orange lines represent the optical paths from different inputs. **d.** Measured input-output transmission map for the various system configurations, where each row indicates the output from all input ports. The configurations include **i**) the initial state,

where the random phase error causes it to deviate from the ideal $O_{4231}$ configuration, and configurations with output permutation of **ii)** $O_{4231}$, **iii)** $O_{4213}$, **iv)** $O_{2413}$, **v)** $O_{1423}$, **vi)** $O_{1324}$. The plotted transmission is averaged from 1335 to 1345 nm and then normalized to the highest total transmission among four inputs.

We measured the initial state of the MZI meshes by using a single tunable-wavelength laser to inject light into each input port and measure four output ports with a power meter (see Methods for details). Repeating this process for all four input ports, we obtained the 4 × 4 in-out mapping matrix (Figure 2d). We present the measured input-output mapping for the initial configuration in Figure 2c(i), where each row shows the output intensity across four output ports if laser is from a particular input port. The map is an averaged transmission from 1335 to 1345 nm wavelength range and normalized to the row with the highest total output power. This mapping is close to the ideal $O_{4231}$ permutation (all-cross configuration shown in Figure 2b) but exhibits low transmission and large crosstalk. This is attributed to initial random phase errors due to fabrication imperfections, especially nonuniform PCM thickness.

To create a different configuration of optical switching fabric, we first identified the minimal voltage to trigger the phase transition of each $Sb_2Se_3$ phase shifter. The individual switching condition tests are necessary since each phase shifter has slightly different electrical resistance (Supplementary Section 9) due to non-ideal metal wire routing. Besides, the structural difference beneath the metal wires can also cause variations in heat dissipation, leading to different thermal behaviors. In the future, the metal routing and the underneath photonic devices can be electro-thermally co-optimized to equalize the resistance and heat dissipation, hence simplifying the condition tests. We first found the amorphization condition by gradually increasing the pulse voltage with a fixed duration of 400 ns until the readout changed, which gives the minimum amorphization voltage. After that, we applied pulses with a 5-ms duration and a slow voltage ramp-up to estimate the crystallization conditions. Supplementary Table 1 shows the detailed switching voltages and pulse durations for all components in the rectangular MZI meshes. These voltage conditions were later used as the minimum phase transition point, based on which incremental voltage renders quasi-continuous tuning.

To configure the system from the initial state to a specified configuration, the intended optical paths were first identified, defined as routes of light that are changed after the reconfiguration. As an example, Figure 2b and Figure 2c show how the OCS is reconfigured from $O_{4231}$ into $O_{4213}$ (shown by dashed paths in Figure 2c). Once an intended path was identified, we adjusted each component to maximize the desired output signal on the intended path. Here, we monitored $I_2O_2$ and tuned M5 to increase its transmitted power at 1340 nm from -40 dBm to -21.2 dBm. In addition to maximizing the desired output, we also minimized crosstalk to other output ports. For example, in the same configuration, the original path $I_3O_3$ is a crosstalk channel, and we minimized its transmission by tuning M5. By selectively monitoring all the output ports, we progressively tuned all components to their correct states. Here we again utilize our bi-directional and "program-and-verify" tuning method. Although the reversible switching has a much smaller range (~$0.23\pi$) compared to the large phase shift of ~$\pi$ required, it was enough for the precise optimization. Note again that this issue of reduced contrast is not fundamental for PCMs, but rather a consequence of our suboptimal material deposition in this particular experiment. Instead of solely monitoring the power at the output gratings, on-chip waveguide taps[50] or transparent in-circuit detectors[51] would allow simultaneous access to all optical powers at each intermediate and output spots, eliminating the need for path-by-path analysis. This would enable a progressive algorithm to reduce crosstalk across all channels, and hence a fully self-reconfigurable system.

With this "program-and-verify" method discussed above, we then demonstrate the various configurations of the OCS based on the circulating MZI meshes. Starting from the initial configuration, we achieved five different system configurations in Figure 2d(ii)–(vi), for which the complete programming sequence are detailed in Supplementary Table 3. The transmission map shows a nearly negligible crosstalk (~ -25 dB). Although a non-ideal loss occurs in Figure 2d(vi), this loss is not from the PCM phase shifter or other silicon photonic components. The PCM phase shifter has a much lower loss of only 0.03 dB/$\pi$. Besides, in the next section, we demonstrate ring resonators using the same system, showing intrinsic Q of more than 226,000 establishing extremely low loss from all components (waveguides, bendings, MMIs, and $Sb_2Se_3$ phase shifters). We attributed the loss in Figure 2c(vi) to undesired crosstalk circling back to the input ports, which could not be captured in our current feedback loop. However, this can be optimized by using fiber arrays and circulators in the ports to allow full monitoring of the system output for feedback.

# Circulating Mach-Zehnder Interferometer mesh structure for coupled microring resonators

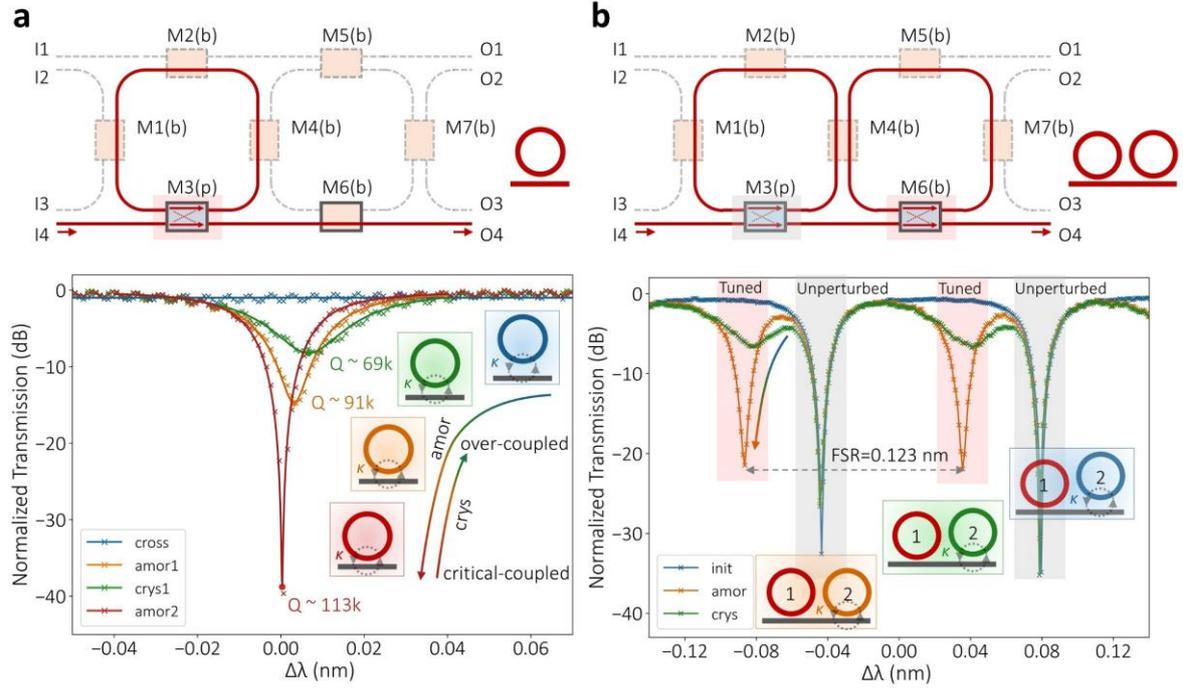

Figure 3: **Nonvolatile MZI mesh realization of single and dual micro-ring resonators**. **a.** Schematic (top) and measured spectra (bottom) for the implemented micro-ring resonator. The schematic shows the states of each MZI unit, where dashed lines with light gray color are MZI units seeing no light. Different colored spectral plots correspond to different coupling ratios, realized by tuning M3 (boxed in red). The crossing markers show the original experimental data (solid lines are Lorentzian fit). The resonance wavelength is indicated by the dot. The quality factors for three different cases are annotated, suggesting a high intrinsic quality factor of around 113,000 (red). The coupling condition is tuned from over-coupled (orange) to critically coupled (red) by applying an amorphization pulse (pulse amplitude 8.8 V, duration 400 ns) and brought back to over-coupled (green) by a crystallization pulse (amplitude 2.5 V, duration 5 ms). **b**. Schematic (top) and measured spectra (bottom) for dual decoupled micro-rings sharing the same bus waveguide. The first micro-ring was fixed (M3 is unchanged indicated by the gray box) while the coupling rate of the second one was changed by tuning M7 (boxed in red). The free-spectral range (FSR) of the resonances are around 0.123 nm, dictated by the long round-trip length of ~ 4 mm for the rectangular meshes. The pulse conditions for M7 were 8.2 V, 400 ns (amorphization) and 2.5 V, 5 ms (crystallization). By tuning M7, a second resonance emerged and was reconfigured. By applying five amorphization pulses, the second ring shows an extinction ratio of > 20 dB (orange), indicating a near-critically coupled behavior. After a crystallization pulse, the second micro-ring was reset to the over-coupled state (green). Thanks to the threshold tuning of PCMs, two micro-rings were tuned in completely thermal crosstalk-free manner, indicated by the unperturbed spectra under the gray box. Amor: amorphization; Crys: crystallization.

To implement a resonant system using the rectangular MZI meshes, we first programmed the mesh to the permutation $O_{1423}$, the same as Figure 2c(v), and then set M4 and M1 to full bar state by minimizing $I_3O_2$ and maximizing $I_4O_4$, respectively. This established a drastically over-coupled microring resonator with a ~100% coupling ratio to the bus waveguide, whose spectrum is the blue line in Figure 3a. By fine tuning M3 and simultaneously monitoring the spectrum, we obtained different waveguide-microring coupling conditions ranging from over coupling to critical coupling in Figure 3a. The blue shift in the spectrum is due to the optical phase modulation of the MZI besides the coupling ratio. Note the orange line is obtained by applying a partial crystallization pulse on the critically coupled system, showing the bidirectional switching capability. We fit the measured data using a Lorentzian lineshape, obtaining quality-factors (Q-factors) of ~69K, ~91K and ~113K for three different coupling coefficients. The high Q-factor in the critically coupled microring suggests a high intrinsic Q-factor of ~226K, establishing a low round trip loss in the resonator (despite the presence of the PCM). The free-spectral range (FSR) is ~0.123 nm, which is mainly limited by the large round-trip propagation distance of ~2 mm, and can be improved significantly by designing more compact photonic components or by adding photonic defects (meshes with different sizes) in the circulating mesh[52].

Next, we demonstrate independent resonance tuning of two decoupled microring resonators sharing a common bus waveguide. Based on the critically coupled microring system configuration in Figure 3a, we first set M6 to the full-cross state by minimizing $I_4O_4$ and M7 to the full-bar state by minimizing $I_4O_2$. Then we tuned M6 for a proper coupling ratio to give the 2$^{nd}$ resonance in Figure 3b (boxed in red). Again, a reversible tuning from a high extinction ratio resonance (orange line) to a lower one (green line) is shown. We highlight that the resonance of Ring 1 is unperturbed while programming Ring 2, validating the thermal crosstalk immunity of PCMs.

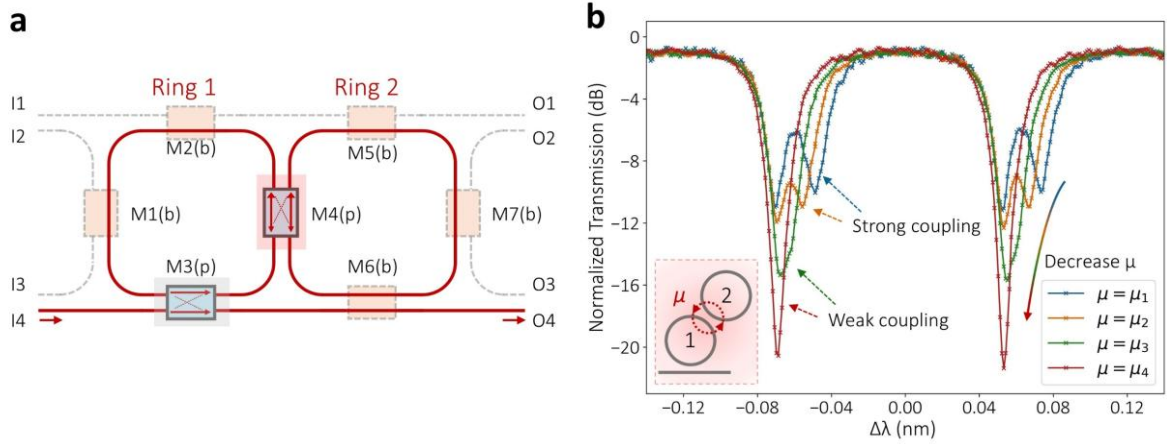

Figure 4: **Implementation of coupled ring resonators with tunable coupling rate. a.** Schematic of coupled micro-ring resonators with two identical rings, with only Ring 1 coupled to the bus waveguide. By controlling M4 we can control the coupling strength between two resonators. **b.** Measured optical spectra of the coupled ring resonator. We observe controlled transition from strong coupling (signified by two dips) to weak coupling (one resonance dip) regime as we decrease the coupling rate $\mu$ by programming M4.

This ability to precisely and locally tune the optical phase can readily be leveraged to study physics of cavity-cavity interactions, paving the way for topological photonics study using coupled waveguide[53] or cavity[54] arrays. As an example, we realized a coupled-ring resonator system with two identical rings, where only one ring is coupled to the bus waveguide, as shown in Figure 4a. To experimentally realize such a system, we decoupled Ring 2 from the bus waveguide by tuning M6 to the full-bar state by observing the disappearance of its resonance. This indicates a negligible coupling rate ($\gamma_c$) compared to the intrinsic decay rate ($\gamma_i$) of Ring 2 ($\gamma_c \ll \gamma_i$).

Physics in this coupled ring system is thoroughly discussed in Supplementary Section 12 by an eigenmode and a temporal coupled mode theory (TCMT) analysis. When two ring resonators are resonant to each other, then by controlling the mutual coupling rate $\mu$ between two rings we can drive the coupled system from a weak to strong coupling regime. In the strong coupling regime, signified by the coupling rate being larger than the loss rates, we observe two dips coming from the hybridized supermodes. On the other hand, for low coupling rate, two modes overlap, and only a single resonance is visible.

In experiment, we have Ring 1 coupled to the bus waveguide, and Ring 2 is decoupled from the bus waveguide, to ensure two rings are coupled only via the MZI M4. We decouple the Ring 2 from the bus waveguide by programming M6. To observe the strong coupling, we need to first realize a small detuning $\Delta$ between two resonances. Apart from some inherent detuning due to fabrication, there is additional detuning between the rings as the Ring 2 is decoupled from the bus waveguide. We gradually tune the resonance of the Ring 2 by amorphizing both arms of M5 and M7. We then tune the mutual coupling rate $\mu$ (by controlling M4). We start with a large value of $\mu$ to observe two resonance dips (blue line in Figure 4b), indicated by two split resonances with a similar extinction ratio. As we gradually decreased the coupling rate, the resonance splitting reduced, and eventually the peaks merged, as presented by the green line in Figure 4b. This merged resonance indicates the cavities are weakly coupled.

We emphasize that controlling the coupling rates in coupled ring resonators require both large and precise local tuning. Such local tuning is beyond the capability of any existing mechanism in integrated photonics, and illustrates the prowess of the close-loop feedback control of PCMs. These experiments show the possibility of simulating physical phenomena using PCM-tunable coupled ring systems, which are robust against thermal crosstalk. For example, one potential future study is realizing different topological and tight-binding Hamiltonians[54,55].

## Self-configurable PIC in a forward Mach-Zehnder Interferometer mesh

In addition to the circulating MZI mesh structure, another typical PIC architecture is the forward-only MZI meshes[12,56 23,578,9,18,58]. We coherently sort orthogonal optical beams from the input ports to a different output port[20,21,23] using PCM-programmable forward-only MZI meshes. Figure 5a shows the schematic of this two-stage self-configurable PIC, which can sort two orthogonal beams (with orthogonal input vectors $\vec{x} = [x_1, x_2, x_3, x_4]^T$) to output port $O_1$ and $O_2$.

We realized such a system in an architecture similar to the well-known Clement architecture[17] for arbitrary unitary transformation (see fabricated PIC in Figure 5b). Only a small portion of MZIs were used to implement the two-stage self-reconfigurable PIC, and other components were set to full-bar state through a progressive algorithm discussed in Supplementary File. Besides the unused components, we also put an extra beam expansion block (Supplementary Section 14) in front of the self-configurable PIC, which expands the single input from laser to a four-element input vector $\vec{x}$ or $\vec{x}'$. $\vec{x}$ and $\vec{x}'$ are orthogonal and created by inputting laser from different ports. Different pairs of $x$ and $x'$ orthogonal input are created by programming the beam expansion block. This block is mandated by our single-input-single-output fiber setup and can be avoided in the future using a fiber array.

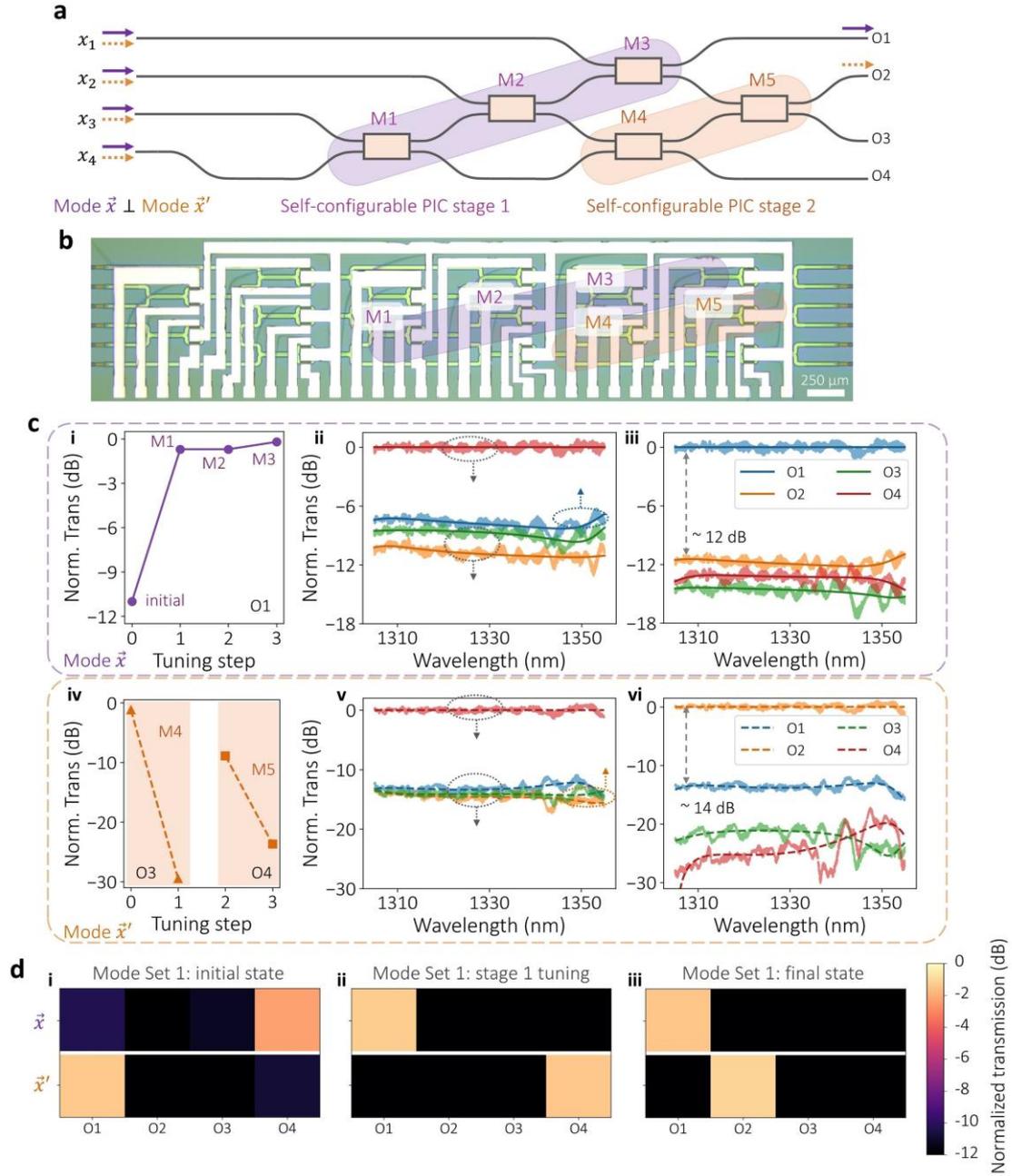

Figure 5: **On-chip self-configurable PIC with non-volatile forward-only MZI mesh. a.** The schematic of the two-stage self-configurable PIC. Four inputs $x_{1\sim4}$ form the input vector $\vec{x}$ of the two stages self-configurable MZIs. Two orthogonal modes $\vec{x}$ and $\vec{x}'$ are sorted to $O_1$ and $O_2$ by programming the first and second stages, respectively. **b.** Optical micrographs of the fabricated forward-only MZI mesh. The MZIs used to implement the self-reconfigurable PIC are annotated corresponding to the schematic. Scalebar: 250 μm. **c**. Beam sorting results for the first pair of orthogonal modes. **i)** The programming process of each MZI in the first stage by maximizing output from $O_1$ at 1340 nm. **ii) iii)** The measured spectra before and after maximizing output from $O_1$, with a crosstalk < -12 dB. **iv)** The programming process of each MZI in the second stage by minimizing output from $O_3$ and $O_4$ at 1340

nm. **v) vi)** The measured spectra before and after minimizing the crosstalk, with a crosstalk of ~ -14 dB. **d.** The transmission map at for two different pairs of orthogonal modes: **i,** for initial state, **ii,** after programming the first stage, **iii,** after programming both stages.

Figure 5c shows an example of the progressive programming of the self-reconfigurable PIC. The first and the second rows correspond to the 1$^{st}$ and the 2$^{nd}$ stages, respectively. We first set an unknown input vector $\vec{x}$. When tuning the first stage, we progressively tuned M1, M2 and M3 to maximize the transmission at the output port $O_1$. Figure 5c(i) shows the initial normalized transmission at $O_1$ and that after programming the first stage, clearly showing an increase of transmission at $O_1$. Figure 5c(ii) shows the initial spectra at four output ports, highlighting a relatively high average crosstalk of ~-8 dB and maximal transmission from output O4. Figure 5c(iii) presents the spectra after optimizing the first stage, exhibiting a maximal transmission of at $O_1$ across a broadband spectrum and a low crosstalk of less than -12 dB. To show the capability of beam sorting, an orthogonal mode $\vec{x}'$ (i.e., $\vec{x}' \perp \vec{x}$) is injected into the self-configurable PIC (see Supplementary Section 14 for implementation). We sorted this orthogonal mode $\vec{x}'$ to the output port $O_2$ without perturbing mode $\vec{x}$, as shown in the second row of Figure 5c. Figure 5c(iv) shows the output at $O_3$ and $O_4$ while programming the MZIs M4 and M5, respectively. These outputs correspond to a crosstalk, which were both reduced to less than -25 dB after optimizing M4 and M5. We note that the minimization method can usually achieve a lower crosstalk because it is less sensitive to intensity fluctuations of the laser. Figure 5c(v) and (vi) present the normalized spectra before and after the optimization, exhibiting a significantly reduced crosstalk. We show the output transmission maps in Figure 5d(i)-(iii), corresponding to the initial state, after programming the first stage and both stages, respectively. The values in the matrices are the normalized transmission average from 1325 to 1330 nm. As labeled in the plot, the first (second) row shows the output transmission with the input vector $\vec{x}$ ($\vec{x}'$), where $\vec{x} \perp \vec{x}'$. From the output map, we verify that the beam sorting of the two self-reconfigurable PIC stages are independent as the first row in Figure 5d(ii) and Figure 5d(iii) are identical.

To further validate the capability of our system, we used another set of two orthogonal beams, as shown in Supplementary Section 14. A crosstalk of less than -10 dB is achieved for the first mode $\vec{x}$, but there is a crosstalk of ~ -5 dB at $O_4$ port for the second mode $\vec{x}'$, significantly more pronounced than Figure 5d(i)-(iii). We attribute this larger crosstalk to the difficulty of independent phase control in our current system. Although it is possible to tune both arms of an MZI to achieve phase-only modulation, decoupling the phase between two outputs and their splitting ratio is impractical in our experiment. Our simulation shows that the lack of independent phase control is likely to cause a large crosstalk (Supplementary Section 15). In the future, this can be further overcome by adding an external phase shifter to our current programmable units[23] (Supplementary Section 15).

## Discussion

We successfully demonstrated the low-loss and precise tuning of high volume manufactured PICs using PCMs, showing a clear path towards a scalable PGA. However, the amount of reversible phase shift in our experiment is only ~$0.23\pi$, which is insufficient to tune the MZIs between their full-bar and full-cross state. Some of our current tuning relies on high contrast switching of PCMs ($> 2\pi$), which led to a gradually reduced contrast (Supplementary Section 16). Such a reduced contrast for large phase tuning is not fundamental and can be avoided by optimizing the material deposition and processing recipe. For example, our previous work with a different $Sb_2Se_3$ source target and deposited with thermal evaporation does not have this issue[46]. Since process variation can lead to limited achievable phase shift, target phase shift of more than $2\pi$ should be designed to add more redundancy and tolerance.

One advantage of PCMs for programmable PICs is its rapid-prototyping nature. Thanks to the nonvolatile nature and crosstalk immunity, PCM phase shifters can be sequentially tuned without interference from previous ones. Since no simultaneous bias source is required to maintain the configuration, the PIC can be set for full functionality without wire bonding or packaging, which can significantly accelerate photonic education and even the prototyping cycle in general. Moreover, the control circuit for this system can be much simpler compared to thermo-optic system. Instead of requiring a digital-analog converter (DAC) for every thermo-optic phase shifter, we can time multiplex the channel addressing process, significantly reducing the number of transistors and multiplexers.

In summary, we demonstrated two classes of MZI-based PCM-programmable PIC. These include a circulating, multipurpose MZI circuit that supports both broadband and resonant functionalities and a forward MZI mesh that can function as an integrated self-configurable MZI system. Although heavily studied using thermo-optic tuning, PCM tuning was not demonstrated in these systems because of the widely spread belief of its unreliability in high-accuracy tuning. Our empirical demonstrations refuted these common beliefs. By using simple feedback provided by external power monitors in a closed-loop tuning, the PCMs can be tuned via "program-and-verify" method, which ultimately guides the system to a desired state. Since PCMs have threshold-triggered tuning behavior, thermal crosstalk is negligible in such large systems and any subsequent tuning will not affect the previous settings.

# Methods

*SOI Device Fabrication*

The initial silicon photonic chips were fabricated in a 300 mm semiconductor fab by Intel Corporation, incorporating standard processes including silicon patterning, doping, and thermal oxide growth. The processed wafers were subsequently diced into 2.5 × 3.3 cm$^2$ reticles for in-house PCM integration. The process for SiO$_2$ window opening and metallization followed a previously reported process[43], with further optimization introduced in this work. Specifically, an additional spin-coating of AZ1512 photoresist, followed by direct-write lithography (DWL) and buffered oxide etching (BOE), was implemented before metallization. The newly added resist layer effectively protected trench sidewalls during BOE, mitigating undercutting. Sb$_2$Se$_3$ deposition windows were defined using 100-kV electron-beam lithography (JEOL JBX-6300FS) on a bilayer resist stack of P(MMA-MAA) copolymer and PMMA to facilitate clean liftoff. A 40-nm-thick Sb$_2$Se$_3$ film (Plasmaterials Ltd.) was deposited by RF magnetron sputtering (Kurt J. Lesker Lab 18). Due to aspect ratio induced shadowing in narrow trenches, the effective film thickness after liftoff was reduced to <10 nm. To encapsulate the PCM, a 40-nm-thick Al$_2$O$_3$ layer was deposited via thermal atomic layer deposition (ALD) at 100 °C (Oxford Plasmalab 80PLUS OpAL). Contact windows to metal pads were opened by an aligned DWL step (AZ1512) followed by chlorine-based inductively coupled plasma reactive ion etching (Oxford PlasmaLab 100 ICP-18). In cases where Sb$_2$Se$_3$ liftoff failed, a 30-second immersion in Piranha solution (3:1 H$_2$SO$_4$:H$_2$O$_2$) was used to selectively remove residual PCM without affecting the underlying silicon or SiO$_2$.

*Optical simulation*

We used Ansys Lumerical INTERCONNECT simulator combined with Python to simulate the self-configurable MZIs, which can be found at *https://github.com/charey6/self-configurable-PIC-simulation*. The Sb$_2$Se$_3$-based silicon phase shifter is designed in Ansys Lumerical finite-difference eigenmode (FDE) with ellipsometry measured Sb$_2$Se$_3$ refractive index data.

*Optical transmission measurement setup*

Optical measurements of $Sb_2Se_3$-based PICs were conducted using a vertically coupled fiber setup with a 20°-angled configuration to minimize back-reflections[43]. The sample stage temperature was stabilized at 26 °C using a thermoelectric controller (TE Technology TC-720) to suppress thermal drifts in temperature-sensitive experiments, such as the implemented microring resonator function. A tunable continuous-wave laser (Santec TSL-510) provided the input light, with polarization adjusted using a manual fiber polarization controller (Thorlabs FPC526) to optimize fiber-to-chip coupling efficiency. Static optical transmission was recorded using a low-noise photodetector module (Keysight 81634B). The spectra were achieved by a LabView program, which continuously records the power when laser wavelength sweeps. Electrical switching of the on-chip phase-change elements was achieved by applying voltage pulses to metal contact pads via two micropositioned electrical probes (Cascade Microtech DPP105-M-AI-S). Crystallization and amorphization pulses were generated by an arbitrary waveform generator (Keysight 81160A). All instrumentation, including the laser, power meter, thermal controller, source meter, and pulse generator, were automated and synchronized using a custom LabVIEW interface. For the automated level setting experiment, a Python-based feedback control algorithm was developed to determine the appropriate voltage levels for precise state tuning. A detailed description of the algorithm is provided in Supplementary Section 5.

**Data availability:** The data that support the findings of this study are available from the corresponding author upon request.

## Acknowledgments

The authors would like to acknowledge Intel Corp. for fabrication of the silicon photonics wafers and thank John Heck, Harel Frish, Haisheng Rong for useful discussions. We also acknowledge Yi-Siou Huang and Professor Carlos A. Ríos Ocampo from University of Maryland, and Professor Juejun Hu from Massachusetts Institute of Technology for insightful discussions and suggestions. The research is funded by the National Science Foundation (NSF-1640986, NSF-2003509), ONR-YIP Award, DARPA-YFA Award, and Intel. Part of this work was conducted at the Washington Nanofabrication Facility/ Molecular Analysis Facility, a National Nanotechnology Coordinated Infrastructure (NNCI) site at the University of Washington, with partial support from the National Science Foundation via awards NNCI-1542101 and NNCI-2025489. Part of this work was performed at the Stanford Nano Shared Facilities (SNSF), supported by the National Science Foundation under award ECCS-1542152).

**Author contributions:** R.C. and A.M. conceived the project. R.C. simulated and fabricated the $Sb_2Se_3$ programmable PICs, performed optical characterizations and data analysis. A.T. helped with the characterization. J.D. and V.T. helped with the fabrication. J.Y. helped with the

# Supplementary Information

# NEO-PGA: Non-volatile electro-optically programmable gate array

Rui Chen[1,2,*], Andrew Tang[1], Jayita Dutta[1], Virat Tara[1], Julian Ye[3], Zhuoran Fang[1], Arka Majumdar[1,3,*]

[1]Department of Electrical and Computer Engineering, University of Washington, Seattle, WA 98195, USA

[2]Department of Materials Science and Engineering, Massachusetts Institute of Technology, MA 02139, USA

[3]Department of Physics, University of Washington, Seattle, WA 98195, USA

*Email: charey@mit.edu and arka@uw.edu

**This supplementary information includes:**

## Section S1: Fabrication process for Back-end-of-line integration

The fabrication process is similar to our previous work[1], where we demonstrated the integration of $Sb_2S_3$. Nevertheless, the process allows the integration of any external materials available from evaporation or sputtering.

We improved the fabrication by splitting the wet etch for PCM window into two steps, which mitigates the undercut since the second wet etch has resist covering most of the sidewalls. The improved fabrication flow is shown below, where the additional wet etch step is highlighted by a red box.

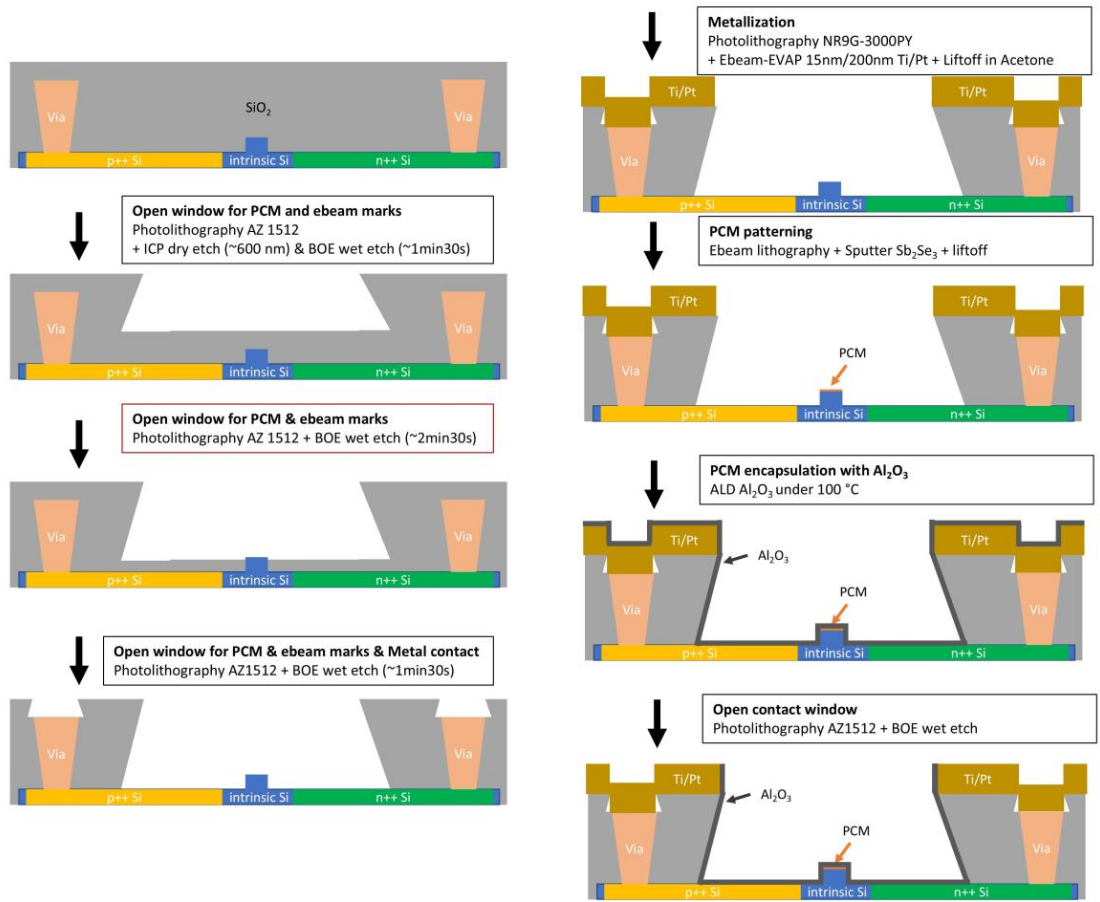

Figure S1: Back-end-of-line integration method for PCM integration.

## Section S2: Complex refractive index of Sb₂Se₃

The refractive index data was measured with ellipsometry and fitted with Tauc-Lorentz model with a mean square error less than 10. The extinction coefficient of both amorphous and crystalline Sb₂Se₃ is close to 0.

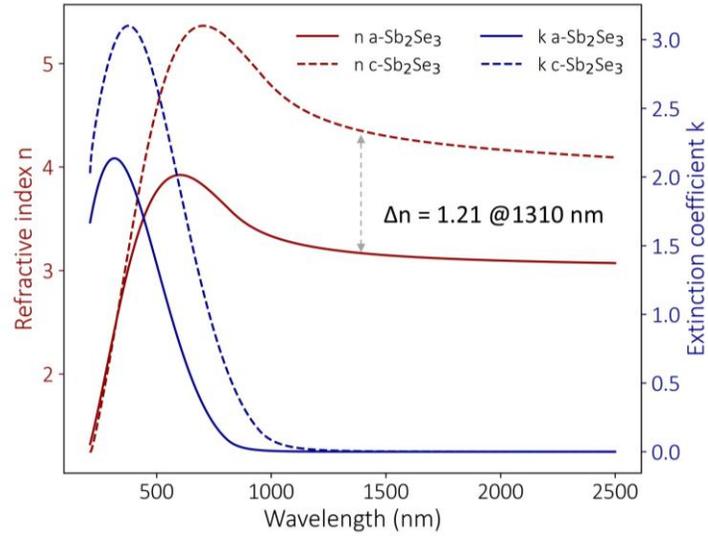

Figure S2: Complex refractive index of sputtered Sb₂Se₃ in both amorphous and crystalline phases.

## Section S3: Micro-ring resonator arrays for phase and loss characterization

The resonance wavelengths for a-Sb$_2$Se$_3$ and c-Sb$_2$Se$_3$ are extracted and analysed, yielding a sampled mean phase shift of $0.0129 \pm 0.00049 \ \pi/\mu m$, or a $\pi$-phase shift of 77.81 μm. The error number for the phase shift estimate is the 95% confidence interval computed from multiple devices with identical Sb$_2$Se$_3$ length.

We note that the loss in the amorphous state is almost negligible, so we focus on the analysis of crystallization loss here. Similar to a previous work[2], we used the following equation to extract the round-trip loss due to PCM:

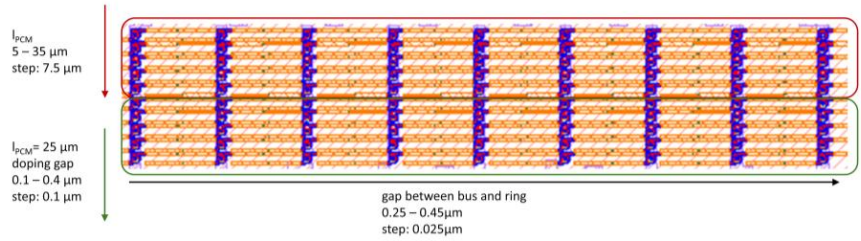

$$Loss(dB) = \frac{10}{\ln(10)} \cdot \frac{2\pi\lambda_0}{FSR} \cdot \left(\frac{1}{Q_c} - \frac{1}{Q_a}\right), \tag{1}$$

where $\lambda_0$ is the resonance wavelength for a-SbSe, FSR is the free-spectral range, $Q_{a,c}$ are the quality factor for a- and c-SbSe respectively. We obtained a loss estimate of $(5.1 \pm 2.9) \times 10^{-4} \ dB/\mu m$ or $0.040 \pm 0.023 \ dB/\pi$, showing less than $0.1 dB/\pi$ level loss, which is highly desired for phase-only systemsW.

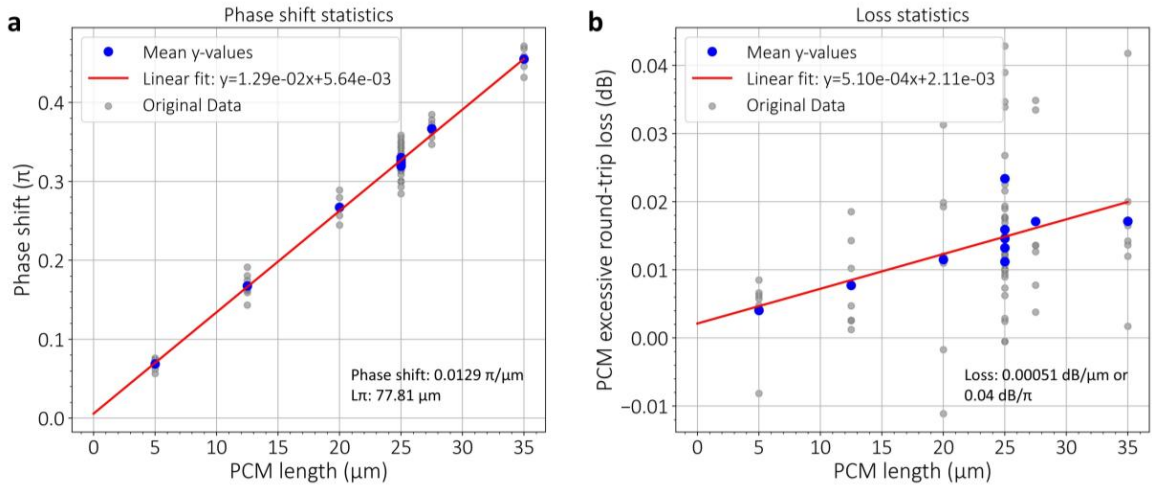

Figure S3: Design of experiment of the micro-ring resonator arrays and an example of the measured ring resonance spectrum and fitting.

Figure S4: Phase shift and loss with respect to different PCM lengths. Extract unit length phase shift and dissipation.

## Section S4: Open-loop multilevel switching of $Sb_2Se_3$ with gradually increased amplitude

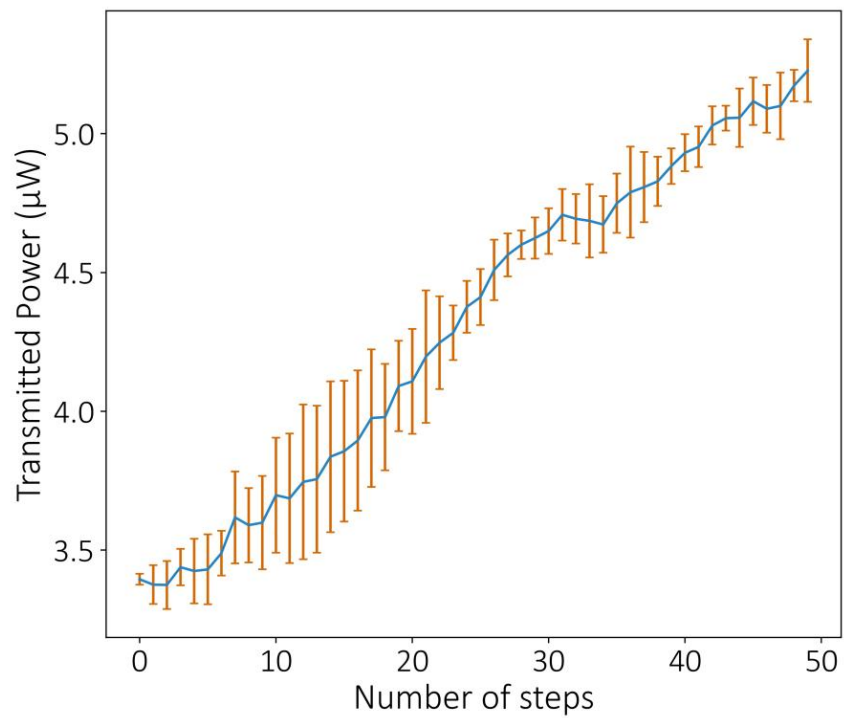

Figure S5: Open-looped multilevel switching results. The amorphization pulses starts from 7.8V, with a step increment of 0.01V and total 50 steps. Although the error represented by standard deviation / mean value is only 3%, it is large enough to have inconsistent level each time. This is to be compared with results in Fig. 1 and Fig. S6, where the desired level is achieved each time through a closed-loop feedback control.

## Section S5: Python implementation of the "program-and-verify" algorithms

   I. Search initial and maximum amorphization and crystallization pulse voltage.
  II. Measure transmission at desired output port and stop the algorithm if it is within threshold of desired power.
 III. If the measured transmission is greater than desired, send amorphization pulse to PCM, or if less than desired send crystallization pulse.
      a. Increase the voltage of the applied pulse if transmission shift is below threshold based off current distance from desired power level.
         i. If the maximum voltage threshold is hit stop the algorithm.
      b. Decrease the voltage of the applied pulse if desired power level is overshot.
  IV. Repeat starting at step II with updated amorphization or crystallization voltage if this loop has not run over the maximum allowed iterations.

**Section S6: More examples of accurate level setting with the feedback**

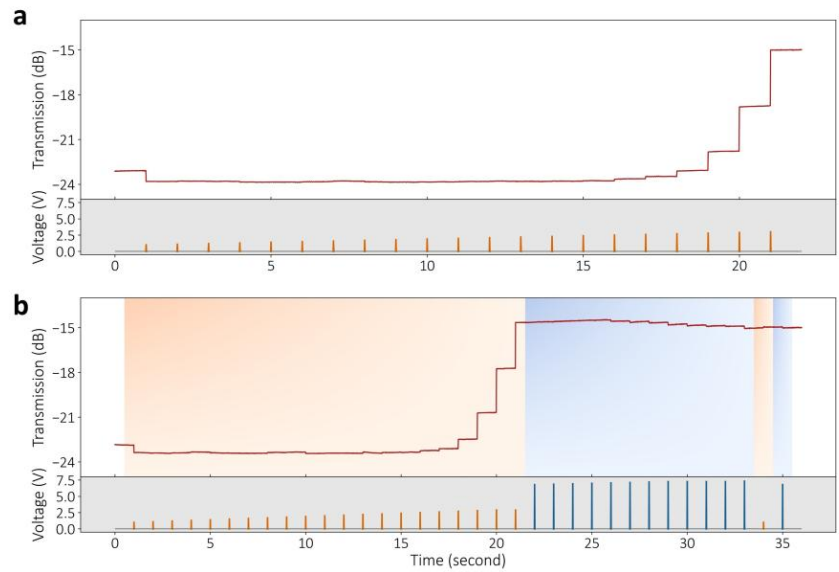

Figure S6: Two more examples. a, example targeting -15 dB with an accuracy of 0.5%; b, the same targeted power level of -15 dB and accuracy of 0.1% as in the main manuscript but exhibits a drastically different tuning path, validating the importance of such program-and-verify method.

**Section S7: Algorithms for maximizing/minimizing optical power**

  I. Define initial and maximum amorphization and crystallization pulse voltage, and set initial pulse to amorphization
 II. Measure transmission at desired output port
     a. Stop the algorithm if it is within threshold of max power
III. Send pulse
     a. Increase the voltage of the applied pulse if transmission shift is below threshold based off current distance to desired power level unless it is already at the maximum allowed voltage
         i. Stop the algorithm if the maximum voltage is hit and increase in transmission is below set threshold
     b. If pulse decreases power switch pulse type and increase voltage of new pulse
 IV. Repeat starting at step II with updated pulse/voltage conditions if this loop has not run over the maximum allowed iterations

During the experiment, we tune the MZIs either to maximize the intended signal or to minimize the undesired crosstalk. The intensity fluctuation at the power meter led to less accurate programming when we maximized the intended signal. As a simplified example, in a two port system, fluctuation of signal between -0.1 dB and -1 dB (typical fluctuation level in the experiment) yields a crosstalk ranging from -16.4 dB to -6.9 dB. This analysis suggests that a pure maximization method cannot eliminate the crosstalk completely. On the other hand, when we minimized the crosstalk, the power meter readout can be down to -30 dB with fluctuation of a few dBs. This much smaller crosstalk uncertainty indicates the minimization of crosstalk method is more accurate and preferred. However, due to the lack of in-circuit monitors, we had to take the maximization approach to optimize some intended optical paths. To overcome this accuracy limitation of the maximization method, we could reduce the intensity fluctuation by using a source with a smaller coherent phase error, such as partial coherent light from an EDFA[10] or μLEDs. Another potential approach is to set all the MZIs after the first stage to bar state. This makes all the later MZIs effectively waveguides, allowing us to use the minimization method without in-circuit monitors. To this regard, we propose an algorithm to set the entire system to an all-bar state by using a progressive method with closed-loop feedback (see Supplementary Slides).

## Section S8: Discussion about the OCS implemented by the 1 × 2 rectangular meshes

Usually, non-blocking switch fabric is desired, where the signal can be arbitrarily guided to any output ports. We first show first that this 1 × 2 rectangular MZ meshes cannot be non-blocking if we only consider full-bar and full-cross state. This is because the vertical MZIs cannot be in the full-bar state, which causes the input to be guided back to another input port (see Figure as an example). Note it is also not possible to have multiple of the vertical MZIs set to bar states without having light circulating back. If M1 is set to bar state, I2 always goes to I3 no matter other parts of the PIC. Similarly, if M7 is set to bar state, O2 and O3 will never be able to have light output. The horizontal MZIs can be set to arbitrary states, so the total possible number of system configuration is $2^4 = 16$. Since the total amount of configuration for a non-blocking switch fabric is $4! = 24$, our system cannot achieve all possible routes and is blocking.

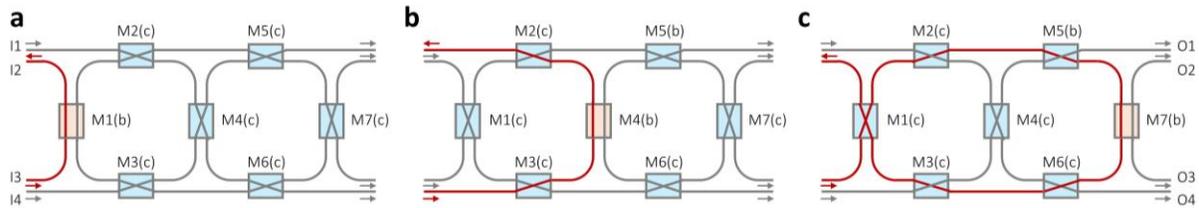

Figure S7: Schematic showing the optical path when the vertical MZIs are set to bar state. The output on the left side is not desired in OCS applications, therefore the vertical MZIs should never be set to bar ports in this mesh design. **a – c**. M1, M4, M7 is set to bar state respectively.

We listed all the possible system permutations as below:

1243, 1324, 1342, 1423; 2134, 2314, 2413, 2431; 3124, 3142, 3241, 3421; 4132, 4231, 4213, 4312

Therefore, the missing permutations are:

1234, 1432; 2143, 2341; 3214, 3412; 4123, 4321

One can find that the missing ones can all be constructed by acting the permutation $O_{1243}$ on accessible permutations. This fact indicates that cascading two such rectangular MZ meshes horizontally to form a 1 × 4 will provide a non-blocking switch fabric.

One interesting problem to investigate further is whether such constrain on the vertical MZIs always holds for larger size of MZI meshes. Firstly, for any $1 \times N$ mesh, this constraint is always true, because the light always bends back to the first vertical MZI and there is no path for it to return to the right side. The case is not as clear for mesh with multiple rows. We leave this problem to later research.

## Section S9: Resistance variation of PIN diodes

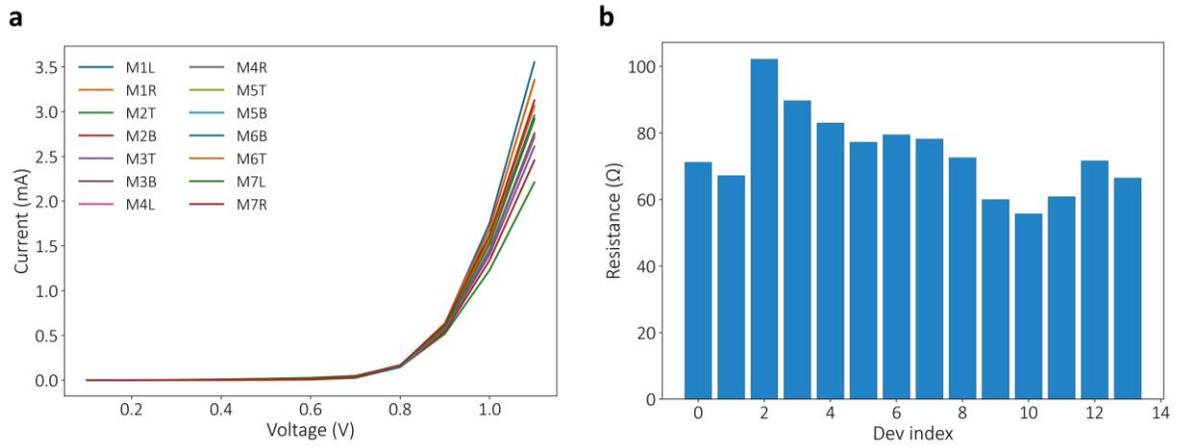

Figure S8: Measured IV curves and the extracted resistance for different PIN diodes in the 1 × 2 rectangular meshes.

## Section S10: Switching conditions

Table S1: Switching conditions for all MZIs in the circulating rectangular meshes.

|    | Arm | I/O  | $V_a$ (V) | $T_c$ ($T_a$) (dBm) | $V_c$ (V) | $T_a$ ($T_c$) (dBm) |
|----|-----|------|-----------|---------------------|-----------|---------------------|
| **M1** | L | I3O3 | 8.3 | -28.6 (-26.6) | 3.3 | -26.6 (-27.7) |
|    | R | I3O3 | 8.4 | -27.8 (-30.0) | 3.3 | -30.3 (-28.1) |
| **M2** | T | I1O4 | 9.7 | -20.8 (-22.9) | 3.6 | -22.9 (-21.8) |
|    | B | I1O4 | 9.5 | -21.8 (-20.2) | 3.6 | -20.2 (-21.1) |
| **M3** | T | I4O1 | 8.6 | -22.4 (-24.8) | 3.1 | -24.8 (-23.0) |
|    | B | I4O1 | 8.5 | -22.8 (-24.0) | 3.2 | -24.0 (-23.2) |
| **M4** | L | I4O1 | 9.2 | -22.3 (-23.5) | 3.1 | -23.5 (-22.0) |
|    | R | I4O1 | 9.5 | -21.7 (-29.2) | 3.6 | -29.2 (-22.6) |
| **M5** | T | I4O1 | 7.8 | -20.7 (-20.0) | 3.1 | -20.0 (-20.3) |
|    | B | I4O1 | 7.8 | -20.3 (-21.3) | 3.1 | -21.3 (-20.8) |
| **M6** | T | I2O2 | 7.4 | -28.5 (-31.2) | 3.0 | -31.2 (-29.0) |
|    | B | I2O2 | 7.5 | -29.0 (-30.5) | 2.9 | -30.5 (-28.9) |
| **M7** | L | I3O3 | 7.9 | -32.5 (-30.2) | 3.2 | -30.2 (-31.2) |
|    | R | I3O3 | 8.1 | -31.0 (-32.8) | 3.2 | -32.8 (-31.6) |

Note: for all the amorphization (crystallization) pulses, the pulse has a pulse duration of 400 ns (10 ms), trailing edge of 6 ns (1 ms) and falling edge of 6 ns (5 ms).

Table S2: Switching conditions for all MZIs in the forward-only, Clement-like meshes.

|    | Arm | I/O  | $V_a$ (V) | $T_c$ ($T_a$) (dBm) | $V_c$ (V) | $T_a$ ($T_c$) (dBm) |
|----|-----|------|-----------|---------------------|-----------|---------------------|
| **M1** | T | I5O2 | 9.4 | -33.6 (-35.3) | 3.2 | -35.3 (-33.7) |
|    | B | I5O2 | 9.4 | -33.6 (-34.0) | 3.1 | -34.0 (-32.9) |
| **M2** | T | I5O2 | 8.1 | -33.2 (-32.1) | 3.0 | -32.1 (-33.4) |
|    | B | I5O2 | 7.8 | -33.6 (-35.0) | 3.0 | -35.0 (-34.0) |
| **M3** | T | I5O2 | 9.1 | -34.2 (-34.7) | 3.2 | -34.7 (-33.9) |
|    | B | I5O2 | 9.0 | -33.6 (-35.0) | 3.2 | -35.0 (-33.8) |
| **M4** | T | I5O2 | 9.1 | -34.1 (-35.2) | 3.5 | -35.2 (-34.5) |
|    | B | I5O2 | 9.9 | -34.2 (-35.1) | 3.3 | -35.1 (-34.7) |
| **M5** | T | I6O1 | 7.9 | -23.0 (-25.3) | 2.8 | -25.3 (-23.0) |
|    | B | I6O1 | 7.8 | -23.0 (-26.6) | 2.7 | -26.6 (-23.1) |
| **M6** | T | I6O1 | 9.1 | -23.5 (-24.3) | 3.4 | -24.3 (-23.5) |
| | B | I6O1 | 9.1 | -23.6 (-25.4) | 3.2 | -25.4 (-23.8) |
| **M7** | T | I6O1 | 9.9 | -23.6 (-25.3) | 3.7 | -25.3 (-23.8) |
| | B | I6O1 | 9.9 | -23.8 (-25.1) | 3.5 | -25.1 (-23.8) |
| **M8** | T | I6O1 | 9.4 | -22.2 (-24.6) | 3.3 | -24.6 (-22.3) |
| | B | I6O1 | 9.3 | -22.3 (-23.5) | 3.2 | -23.5 (-22.3) |
| **M9** | T | I6O1 | 8.8 | -22.4 (-21.6) | 3.2 | -21.6 (-22.2) |
| | B | I6O1 | 8.6 | -22.1 (-24.2) | 2.9 | -24.2 (-22.2) |
| **M10** | T | I4O4 | 9.9 | -39.0 (-40.0) | 3.2 | - 40 (-38.9) |
| | B | I4O4 | 9.8 | -38.5 (-39.6) | 3.2 | -39.6 (-38.5) |
| **M11** | T | I4O4 | 8.9 | -38.4 (-36.4) | 3.1 | -36.4 (-37.4) |
| | B | I4O4 | 8.9 | -37.6 (-39.0) | 3.3 | Not that reversible |
| **M12** | T | I4O4 | 8.0 | -38.5 (-39.7) | 3.2 | -39.7 (-38.3) |
| | B | I4O4 | 7.8 | -38.1 (-36.5) | 3.3 | -36.5 (-37.2) |
| **M13** | T | I4O5 | 7.5 | -24.5 (-26.1) | 3.3 | -26.1 (-25.7) |
| | B | I4O5 | 7.1 | -25.6 (-24.7) | 3.0 | Not that reversible |
| **M0** | T | I1O6 | 8.9 | -37.7 (-39.5) | 3.2 | -39.5 (-37.8) |
| | B | I1O6 | 8.9 | -37.2 (-42.2) | 3.2 | -42.2 (-37.5) |

## Section S11: Programming sequence

Table S3: Programming sequence for implementing different OCS configurations with the 1 × 2 circulating meshes.

| Device | I/O | Action: a or c | dT (dBm) |
|---|---|---|---|
| **CONFIG2** | | | |
| **M1L** | I3O3 | a: 8.7V | -28.5 -> -24.0 |
| **M2T** | I3O3 | a: 9.7V | -24.0 -> -24.5 |
| | I3O3 | c: 3.5V | -24.5 -> -24.0 |
| **M2B** | I3O3 | a: 9.7V | -24.0 -> -23.5 |
| **M5T** | I3O3 | a: 8.0V | -23.5 -> -22.0 |
| | I3O3 | c: 2.9V | -22.0 -> -21.7 |
| **M7L** | I3O3 | a: 8.0V | -21.7 -> -21.2 |
| **M6T** | I2O4 | a: 7.4V | -30.0 -> -62 |
| **M5T** | I2O1 | a: 8.6V | -36.0 -> -60 |
| **M3T** | I2O2 | a: 8.6V | -20.5 -> -20.1 |
| | | | |
| **CONFIG1** | | | |
| **M5T** | I4O1 | c: 3.2V | -40.0 -> -26.7 |
| **M5B** | I4O1 | a: 7.8V | -26.7 -> -22.6 |
| | | | |
| **CONFIG2** | | | |
| **M5T** | I4O1 | a: 9.1V | -20.4 -> 59 |
| | | | |
| **CONFIG6** | | | |
| **M6T** | I2O2 | a: 8.1V | -20.6 -> -64 |
| | | | |
| **CONFIG 14** | | | |
| **M2T** | I3O1 | a: 10V, 430 ns | -20.5 -> -56 |
| | | | |
| **CONFIG16** | | | |
| **M3T** | I2O4 | a: 8.8V | -20.6 -> -56 |
| | | | |
| **CONFIG14** | | | |
| **M3B** | I4O4 | a: 8.4V | -19.8 -> -53 |

Table S4: Programming sequence for implementing ring resonators with the 1 × 2 circulating meshes.

| Device | I/O | Action: a or c | dT (dBm) |
|---|---|---|---|
| **CONFIG14 - set M4** | | | |
| **M4L** | I3O2 | a: 9.8V | -20.7 -> -63 |
| | | | |
| **CONFIG14 - set M1** | | | |
| **M1L** | I4O4 | a: 9.0V | -40 -> -22.0 |
| | | | |
| **Single Resonator** | | | |
| **M3B** | I4O4 | a1: 8.7V | Check resonance |
| | | c1: 3.1V | |
| | | | |
| **Dual-ring resonator decoupled** | | | |
| **M6T** | I4O4 | a: 8.4V | -20.1 -> -40 |
| | | c: 2.5V | -40 -> -60 |
| | | | |
| **M7L** | I4O2 | a: 8.2V | -20.8 -> -60 |
| | | | |
| **Dual-ring resonator coupled** | | | |
| **M6T** | I4O4 | c: 3.0V | |
| | | c2: 3.3V | |
| **M6B** | I4O4 | a1: 7.4V | |
| | | c1: 2.9V | |
| **M6T** | I4O4@1350.05nm | a1: 8.7V | |
| **M4L** | I4O4 | c1: 3.1V | |
| **M6B** | I4O4@1350.05nm | a1: 7.6V | over coupled |
| | | a2: more 7.6V | under coupled |
| | | a5 | |
| **M4L** | I4O4 | a1: 9.6V | |
| | | a2: 9.7V | |
| | | a5: 10V | |
| | | a6: 410 ns | |
| | | a9: 440 ns | |
| | | a11: 450 ns | |
| | | a12: 460 ns | |
| | | every one increase by 10 ns | |
| | | c1: 3.3V | |

Table S5: Programming sequence for beam sorting in the self-configurable PICs.

| Device | I/O | Action: a or c | dT (dBm) |
|---|---|---|---|
| **Set bar** | | | |
| **M5T** | I1O6 | a1: 8.0V | -20.1 -> -57 |
| **M10T** | I1O4 | a1: 10.0V | -33.7 -> -65 |
| **M13T** | I1O2 | a1: 8.2V | -22.4 -> -53 |
| **M0T** | I6O2 | a1: 9.0V | -33.2 -> -70 |
| **M4T** | I6O3 | a1: 10V × 20 + c1: 3V | -22.2 -> -58 |
| **M9T** | I6O5 | a1: 9.2V | -20.2 -> -56 |
| **Self-aligning the original modes** | | | |
| **stage 1** | | | |
| **M6T** | I2O5 | a1: 9.1V | -30.4 -> -32.4 |
| | | c1: 3.2V | -32.4 -> -31.0 |
| **M6B** | | a1: 9.2V × 6 + 9.3V | -31.0 -> -21.7 |
| **M7T** | | a1: 10V, 410 ns | -21.7 -> -21.9 |
| | | c1: 3.6V | -21.9 -> -21.7 |
| **M7B** | | a1: 10V, 410 ns | -21.7 -> -21.9 |
| | | c1: 3.5V | -21.9 -> -21.7 |
| **M8T** | | a: 9.4V + 9.5V | -21.7 -> -21.0 -> -22.0 |
| | | c: 3.5V | -22.0 -> -21.4 |
| **M8B** | | a: 9.3V + 9.4V | -21.4 -> -21.0 -> -21.2 |
| | | c: 3.5V | -21.2 -> -21.2 |
| **stage 2** | | | |
| **M11T** | I3O2 | a1: 9.3V | -22.2 -> 50.5 |
| **M12T** | I3O3 | a1: 8.2V | -29.9 -> 44.7 |
| **Self-aligning the unknown modes by perturbing the MZIs with three partial amorphization pulses** | | | |
| **Perturb pulses** | | | |
| **M1T** | I2O5 | a1: 9.5V × 7 | -21.0 -> 22.9 |
| **M2T** | I2O5 | a1: 8.5V × 6, 8.6V × 10 | -22.9 -> 23.4 |
| **M3T** | I2O5 | a1: 9.2V × 3, 9.3V × 7 | -23.4 -> 22.5 |
| | | | |
| **Stage 1** | | | |
| **M6T** | I2O5 | a1: 9.3V | -27.6 -> 22.7 |
| **M7T** | | a1: 9.9V | -22.7 -> 23.4 |
| | | c1: 3.6V | -23.4 -> 22.7 |
| **M7B** | | a1: 9.9V | -22.8 -> 21.4 - 22.1 |
| | | c1: 3.6V | -22.1 -> 21.8 |
| **M8T** | | a1: 9.3V | -21.7 -> 22.2 |
| | | c1: 3.6V | -22.2 -> 22.0 |
| **M8B** | | a1: 9.4V | -22.0 -> 20.8 |
| **M6T** | I2O4 | a1: 9.4V | -31.7 -> 32 |
| **M7T** | | a1: 9.9V | -31.7 -> 31.5 |
| | | c1: 3.6V | -31.5 -> 31.7 |
| **M6T** | I2O5 | a1: 9.3V | -23.5 -> 22.5 - 23.3 |
| | | c1: 3.3V | -23.3 -> 22.6 |
| **M6B** | | a1: 9.6V | -22.6 -> 22.0 - 23.0 |
| | | c1: 3.3V | -23.0 -> 22.0 |
| **M7T** | | a1: 10.0V, 420 ns | -22.3 -> 21.5 |
| | | c1: 3.5V | -21.5 -> 21.8 |
| | | a1: 10.0V, 420 ns | -21.8 -> 22.5 |
| | | c1: 3.5V | -22.5 -> 21.7 |
| **M7B** | | a1: 9.9V | -21.7 -> 21.8 |
| | | c1: 3.5V | -21.8 -> 21.6 |
| **M8T** | | a1: 9.5V | -21.9 -> 22.1 |
| **M8B** | | a1: 9.6V | -22.1 -> 21.7 |
| | | c1: 3.3V | -21.7 -> 21.5 |
| **stage2** | | | |
| **M11T** | I3O2 | a1: 9.6V | -25.5 -> 27.6 |
| **M11B** | | a1: 9.6V | -27.3 -> 25.0 |
| **M11T** | | a2: 10V, 440 ns | -25 -> 28.0 - 25.3 |
| **M11B** | | a2: 9.8V | -25 -> 28.0 - 23.4 |
| **M11T** | | a3: 10V, 600 ns | -23.4 -> 27.6 |
| **M12T** | I3O3 | a1: 8.0V | -32.2 -> 31.9 |
| **M12B** | | a1: 8.2V | -31.9 -> 33.0 |
| | | c1: 2.7V | -28.8 -> 32.7 |

## Section S12: Model for coupled micro-ring resonators

The physics of this coupled ring system is described by the following Hamiltonian:

$$\begin{bmatrix} \omega_1 - i\gamma_1 & \mu \\ \mu & \omega_2 - i\gamma_2 \end{bmatrix}, \tag{2}$$

where $\omega_{1,2}$ and $\gamma_{1,2}$ are the resonance frequencies and the loss rates in ring 1 and ring 2, and $\mu$ is the mutual ring coupling rate. The eigenfrequencies of this Hamiltonian are given by:

$$\omega_\pm = \bar{\omega} - i\bar{\gamma} \pm \sqrt{\mu^2 + \left(\frac{\Delta - i\delta\gamma}{2}\right)^2}, \tag{3}$$

Where $\bar{\omega} = \frac{\omega_1 + \omega_2}{2}$, $\Delta = \frac{\omega_1 - \omega_2}{2}$, $\bar{\gamma} = \frac{\gamma_1 + \gamma_2}{2}$, $\delta\gamma = \frac{\gamma_1 - \gamma_2}{2}$. We assume both microring resonators have the same intrinsic loss $\gamma_0 = \frac{\bar{\omega}}{Q_0} = \frac{1}{Q_0}$, where $Q_0 = 226{,}000$ is the intrinsic Q-factor, and that ring 1 is critically coupled with the bus waveguide, which yields $\gamma_1 = 2\gamma_0, \gamma_2 = \gamma_0$.

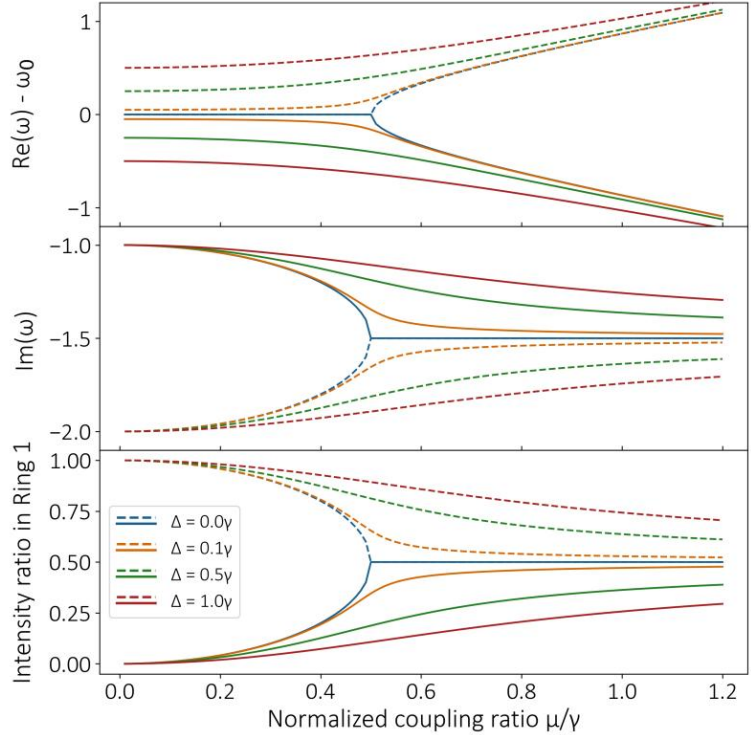

Figure S9: Simulation results from eigenmode analysis using the Hamiltonian of the coupled cavity system consisting of two rings. The results include the real (top), imaginary (middle) part of the eigenfrequencies of the system, and the fraction of mode in ring 1 (bottom). When the detuning $\Delta$ is sufficiently small compared to the intrinsic loss of Ring 1(2), there are two distinct regimes. However, the boundary is blurred as the detuning $\Delta$ increases.

We implemented a Python script to solve the eigenvalues and eigenvectors of the Hamiltonian and obtained Figure S9. The top and middle plots show the real and imaginary part of the eigenfrequencies for different mutual coupling rates, $\mu$, and for various detunings, $\Delta$. An exceptional point (EP) is clearly shown at $\mu = 0.5\gamma$. Specifically in the ideal case of $\Delta = 0$, the coupled resonators exhibit drastically different behavior when $\mu < 0.5\gamma$ and $\mu > 0.5\gamma$, corresponding to the single-dip and dual-dip phase described in the main manuscript. In fact, the bifurcation curve of the complex eigenfrequencies is quite similar to ones in $\mathcal{PT}$-symmetric physics. Below we discuss a little bit about how to link it to a passive $\mathcal{PT}$-symmetric physics picture.

When $\mu > 0.5\gamma$, two resonance frequencies are present, sharing the same loss rate (imaginary part of $\omega_{\pm}$) of $\gamma_0 \approx -\frac{1.5}{\gamma}$. Designating this loss as artificially neutral, we can observe similar behavior to the $\mathcal{PT}$-symmetry physics in conventional loss-gain systems. Since in this phase, the loss of the supermodes is artificially zero in the presence of a loss, it is termed as the passive $\mathcal{PT}$-symmetry phase. When $\mu < 0.5\gamma$, the resonances coalesce and artificial loss and gain emerge in the system, meaning the system reaches $\mathcal{PT}$-symmetry broken phase. Since the system behaves distinctly with $\mu$ greater or smaller than $0.5\gamma$, $\mu = 0.5\gamma$ is called an exceptional point (EP).

We simulated the system for different detunings $\Delta$ and found that $\Delta$ significantly influences the behavior of the system, and that a larger $\Delta$ than $\gamma$ can lead to the disappearance of the EP. Therefore, it is necessary to make sure that both resonators have identical resonance wavelength to observe the two distinct regimes.

Intuitively, we should only see one mode of Ring 1 if $\mu = 0$ (meaning Ring 2 is fully decoupled to the system) regardless of the detuning $\Delta$. But according to the top two plots, when $\Delta$ is relatively large, there exist two resonance modes when $\mu = 0$. We answer this question by providing the last plot of Figure S9(bottom), which shows the mode overlap between two supermodes and each ring. This is obtained by evaluating the eigenvectors. Clearly, one of the two supermode (solid lines) has a smaller mode overlap with Ring 1, hence less coupling to the reservoir (bus waveguide). When $\mu = 0$, one mode is completely dark, which does not couple to the bus waveguide, leading to only one observable resonance.

In addition, we also directly implement the temporal coupled mode theory to study the spectra for different mutual coupling rates, $\mu$, and detunings, $\Delta$. We followed a standard formulation for coupled ring resonators[3] and the formulation is as below:

$$\begin{cases} \frac{da_1}{dt} = (i\omega_1 - \gamma)a_1 - i\mu a_2 + \sqrt{2\gamma_c}s_{in} \\ \frac{da_2}{dt} = (i\omega_2 - \gamma)a_2 - i\mu a_1 \\ s_{out} = s_{in} - \sqrt{2\gamma_c}a_1 \end{cases}, \quad (4)$$

Where $a_{1,2}$ is the electric field amplitude in Ring 1 and 2 respectively, $\gamma$ is the intrinsic loss rate in Ring 1 and 2, $\omega_{1,2}$ is the resonance frequencies in Ring 1 and 2, $\gamma_c$ is the coupling rate between Ring 1 and the bus waveguide, $\mu$ is the mutual coupling rate between Ring 1 and 2. $s_{in}$ and $s_{out}$ are the input and output amplitude. Since no nonlinearity is considered in the system, all the amplitudes have the same frequency as the input, $\omega$. In the steady state, all the differential terms with respect to time $t$ are set to 0's, one can obtain that:

$$a_2 = -i\frac{\mu}{i(\omega-\omega_2)+\gamma}a_1 \tag{5}$$

$$a_1 = \frac{\sqrt{2\gamma_c}s_{in}}{i(\omega-\omega_1)+\gamma+\frac{\mu^2}{i(\omega-\omega_2)+\gamma}} \tag{6}$$

And the final output transmission yields:

$$T(\omega) = \left|\frac{s_{out}}{s_{in}}\right|^2 = \left|1-\frac{2\gamma_c}{i(\omega-\omega_1)+\gamma+\frac{\mu^2}{i(\omega-\omega_2)+\gamma}}\right|^2 \tag{7}$$

The transmission spectrum dictated by Eq. (7) is plotted below in Figure S10 for two cases where detuning between two rings $\Delta = \omega_2 - \omega_1$ is $3\gamma$ or 0. Notably, when $\Delta$ is non-zero, the appearance of the second resonance dip at a large $\mu$ is a result of the increased mode overlap, an entirely different physics from the $\mathcal{PT}$-symmetry-like behavior. We also note that in this simulation we did not consider the shift of $\omega_1$ and $\omega_2$ due to an extra phase shift accompanied by the coupling during our experiment. Therefore, in experiment we observed an extra shift of resonance when $\Delta = 0$.

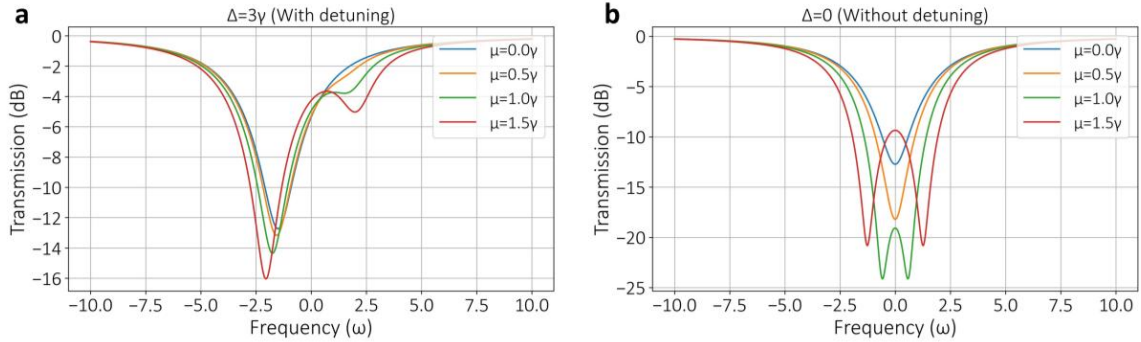

Figure S10: Temporal coupled mode theory simulation of the coupled-ring resonator when the two ring resonators have (a) different and (b) identical resonance wavelengths (or equivalently frequencies). The detuning of the resonance frequencies between two rings $\Delta = \omega_1 - \omega_2$, where $\omega_{1,2}$ are the resonance frequencies of Ring 1 and 2, respectively, and $\omega_1 + \omega_2 = 0$. $\gamma$ is the intrinsic loss of both rings. $\mu$ is the mutual coupling rate between two rings.

# Section S13: Measured results for coupled-cavity system with large detuning

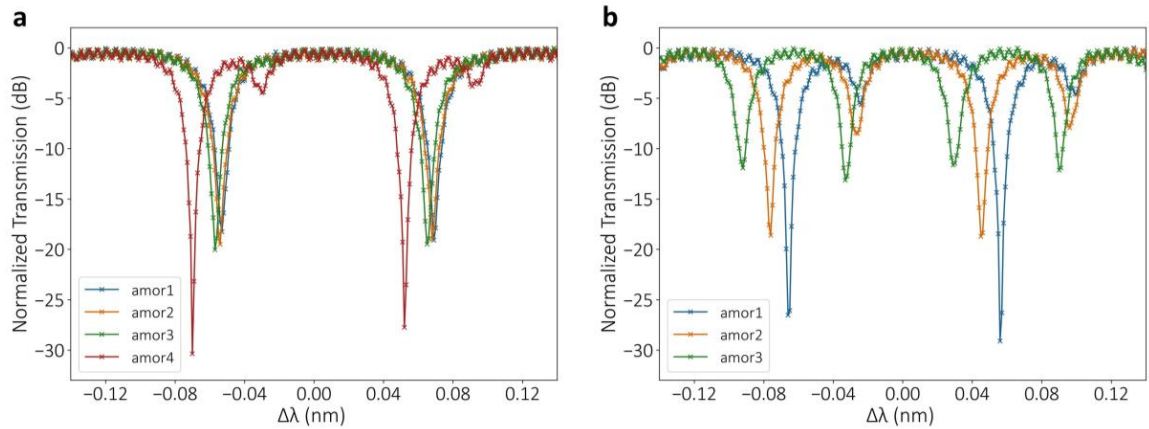

Figure S11: Measured spectra for coupled-ring resonator before aligning the resonance of two rings. (a) Single-dip regime; (b) dual-dip regime. Although the system also evolves from single-dip to dual-dip, the extinction ratio between two dips is different near the "exceptional point" (in fact the system does not possess an EP anymore from the eigenmode analysis). This indicates a non-zero detuning Δ, which is compensated for to produce the results in Fig. 4 in the manuscript.

## Section S14: Full schematic with the beam expansion blocks and creation of orthogonal modes $\vec{x}$ and $\vec{x}'$

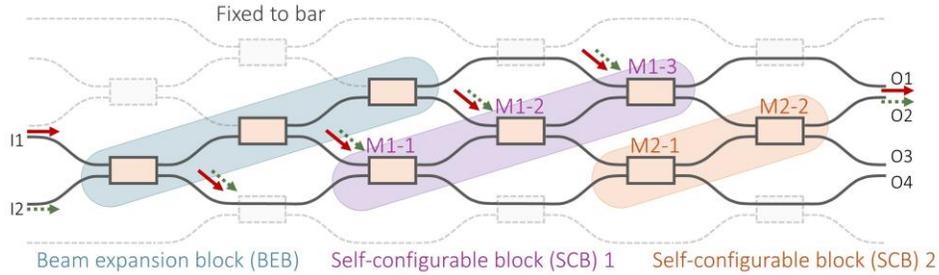

Figure S12: Full schematic of our implementation of the self-reconfigurable PICs, which includes a beam expansion block in front to expand input from a single grating to the vector $\vec{x}$ ($\vec{x}'$), when inputting from I1 (I2).

The new orthogonal mode $\vec{x}'$ is implemented by switching the laser input port from I1 to I2 without changing the beam expansion block, due to: (1) The inputs $[0, 1]^T$ and $[1, 0]^T$ clearly are orthogonal to each other. (2) The PIC effectively establishes a linear transformation $\hat{A}$ on the input modes, which does not change the orthogonality of the input as long as $\hat{A} \cdot \hat{A}^T = \hat{I}$, where $\hat{I}$ is the identity matrix. In the case of unitary transformation (absence of or little loss), this condition is always true in the Clement-like architecture. It is noteworthy that if the system is too lossy, the orthogonality can break down, leading to the creation of nonorthogonal modes[4].

To obtain a totally different orthogonal mode pair $\vec{x}$ and $\vec{x}'$, we program the beam expansion block for a different linear transformation $\hat{A}$. One example is shown below right after we sorted the two beams in Fig. 5. One can see that if the modes $\vec{x}$ and $\vec{x}'$ changed, the self-configurable PIC can be reprogrammed to sort the new beams. The transmission behavior is significantly perturbed in Figure S13 compared with Figure 5, suggested by the non-zero values of almost all elements. Following the same procedure, we obtained the transmission maps in Figure S13(a) and (b) after tuning the first and the second stages, respectively.

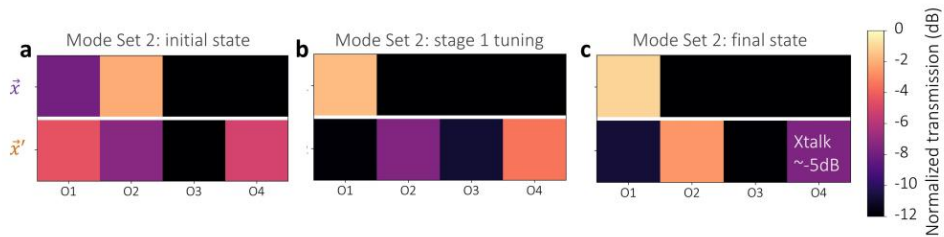

Figure S13: Reprogramming the beam expansion block creates another pair of orthogonal modes $\vec{x}$ and $\vec{x}'$.

## Section S15: Simulation for self-configurable PICs and new unit cells

We used Lumerical Interconnect to simulate the self-configurable PICs, which was controlled by Python (through the *lumapi* interface. We first manually laid out a two-stage self-configurable PIC architecture (Figure S14), in which all the phase (determined by the arm lengths of MZIs) were by default. Note that a tap is used at the crosstalk channel to connect to a monitor ($D_{i,j}$). This was used to sort two orthogonal input vectors of $x = \left[2e^{-i\frac{\pi}{2}}, 2e^{i\frac{\pi}{6}}, 0.25e^{i\frac{\pi}{4}}, e^{i\frac{\pi}{2}}\right]^T$ and $x' = \left[(\frac{7}{4}+\sqrt{3})e^{i\frac{\pi}{2}}, e^{i\frac{\pi}{2}}, 2e^{i\frac{\pi}{4}}, 3e^{i\frac{\pi}{4}}\right]^T$ to two different output ports. Note that this set of orthogonal modes were chosen randomly and we do not see any fundamental issues with using another set of modes.

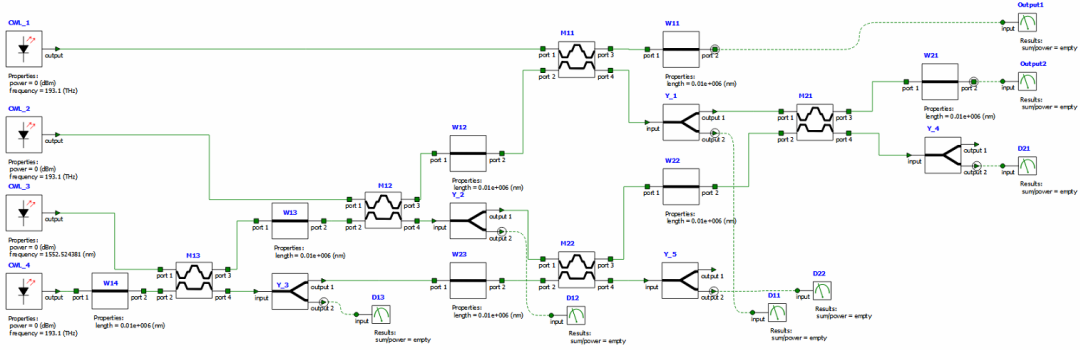

Figure S14: Lumerical Interconnect implementation of self-configurable PIC.

A Python script (see Methods for *github* link) realized the progressive method through a gradient descent algorithm. As a result, output at each undesired monitor was minimized. The gradient $g$ is calculated by a perturbative method. For MZI $M_{i,j}$, where $i$ is the stage index, and $j$ is the component index, two simulations were performed as we perturbed the arm length (effectively the splitting ratio) or the waveguide length (effectively the phase) slightly by $+dl$ and $-dl$, providing a transmission change at monitor $D_{i,j}$ of $\Delta T$. The gradient $g$ was then computed as:

$$g = \frac{\Delta T}{2dl} \quad (8)$$

Then the length $l$ was updated by:

$$l = l - g \cdot r \quad (9)$$

Where $r$ is the updating rate, which determines how fast the algorithm converges. To have a relatively fast and consistent convergence result, we chose $r = 2$.

For each MZI, we ran this algorithm first for the external phase (waveguide length) and then for the splitting ratio (MZI arm length). Since the phase also slightly changed when we changed the arm length of the MZI, we iterate the tuning multiple times (phase, ratio, phase, ratio, …) to compensate for that until it converges. This algorithm then iterated over all the MZIs in order, achieving the desired beam sorting purpose. Below we show the process of optimization, where we can see the phase step $d\phi$ which is proportional to $dl$ decreases with more iterations and the crosstalk reduces. The bumps for $d\phi$ is due to the extra phase to be compensated for when we tuned the splitting ratio of MZIs. As a result, we were able to achieve less than -25 dB crosstalk for all the channels. Note that this number is not fundamentally limited by physics and can certainly be further reduced if we use a smaller update rate $r$ or set a more stringent condition to stop the optimization algorithm.

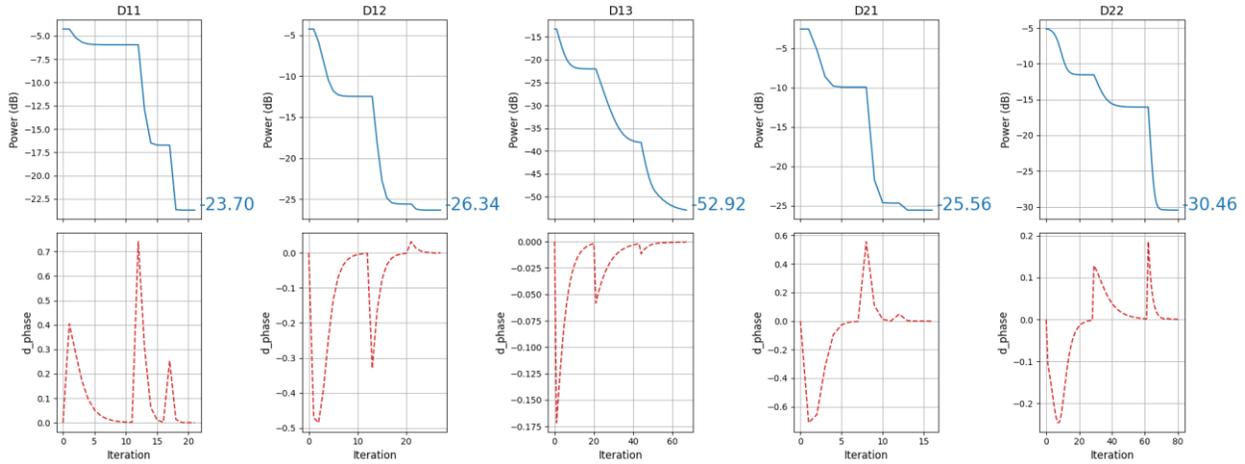

Figure S15: Simulation procedure for self-configurable PIC when tuning both arm lengths in MZIs and waveguide lengths. (Top) The output at each crosstalk monitor. (Bottom) The phase difference $d\phi$ with each iteration. The large $d\phi$ in the middle is because of the extra phase variation when we tune the MZI arm lengths.

Now, we run the same simulation but with only varied MZI arm lengths to mimic our experiment in Figure 5. The results are shown in Figure S18. We stress a relatively large crosstalk with a large crosstalk of -7.8 and -9 dB in D11 and D12. This simulation illustrates that without the external phase shifter the performance of the beam sorter will drop, perfectly matching our experimental observations (the last plot in Figure 5).

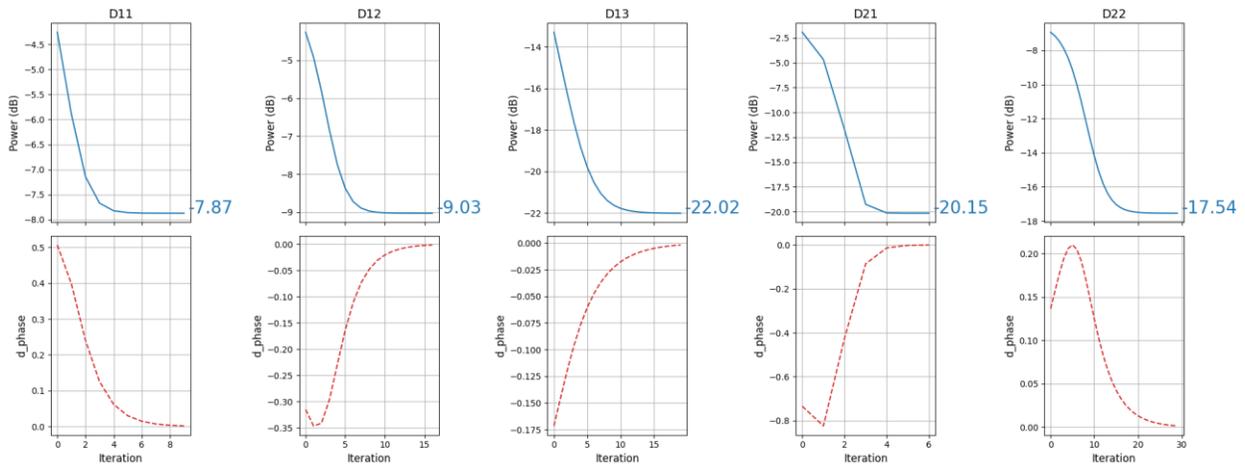

Figure S16: Simulation procedure for self-configurable PIC when only tuning the arm lengths in MZIs. (Top) The output at each crosstalk monitor. (Bottom) The phase difference $d\phi$ with each iteration.

In the future we will use the unit cell shown below to provide both splitting ratio and phase tuning.

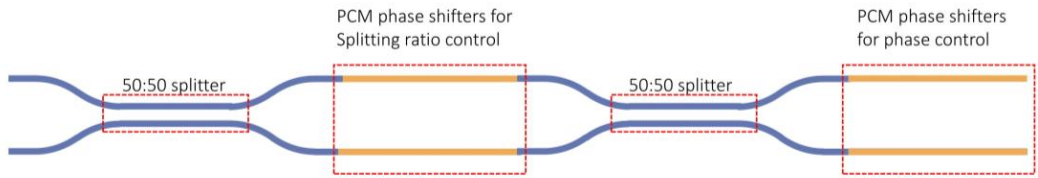

Figure S17: New unit cell with external phase shifters for independent phase control. This is important to achieve low crosstalk in self-configurable PICs.

**Section S16: Gradually reduced contrast for large transmission change**

Our $Sb_2Se_3$ phase shifters suffer from gradual loss of contrast when the required phase shift is large. Besides, we also observed a baseline drift, which was in the same direction as amorphization even when we used a long pulse for crystallization. This can be attributed to the ablation of $Sb_2Se_3$, causing a reduced effective index.

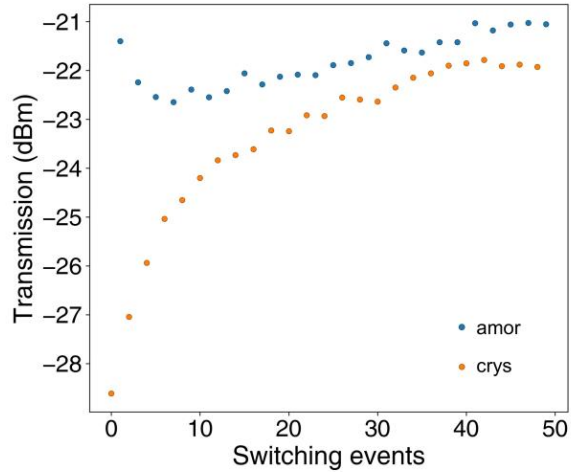

Figure S18: Cyclability test of an $Sb_2Se_3$ phase shifter, illustrating a quick reduction of contrast with more switching events. Although initial contrast is around 8 dB, it gradually reduced to only 1 dB. Besides, a response drift to the blue side is also observed (going up), which corresponds to the amorphization.

However, we note that this can be addressed in the future by having better quality materials. We used a shared sputtering system, which may have led to some material degradation. A recent work of ours shows perfect reversible switching and high endurance using $Sb_2Se_3$ from Prof. Juejun Hu's lab, see Figure 4 of Ref [5]. Although this was done in a university cleanroom with our own silicon-on-insulator chips, we believe this can prove it is possible to achieve fully reversible switching of $Sb_2Se_3$ in the future.